\documentclass[12pt]{article}
\usepackage[utf8]{inputenc}
\usepackage[T1]{fontenc}
\usepackage[USenglish,activeacute]{babel}
\usepackage{varwidth}
\usepackage{cite}
\usepackage{physics}
\usepackage{url}
\usepackage{empheq}
\usepackage{wasysym}
\usepackage{fontawesome}
\usepackage{marvosym}
\usepackage{adforn}
\usepackage{halloweenmath}
\usepackage{staves}
\usepackage{bclogo}
\usepackage{slashed} 
\usepackage{amsthm,amsmath,latexsym,amssymb,amsfonts,amsbsy}
\usepackage{graphicx,lscape,fancyhdr,xcolor,array,stmaryrd,euscript,wrapfig}
\usepackage{caption}
\usepackage{subcaption}
\usepackage{hyperref}
\usepackage{tikz}

\def\lp{\ell_{\rm P}}
\def\cR{{\cal{R}}}
\def\cL{{\cal{L}}}
\def\cS{{\cal{S}}}

\usepackage{tikz}
\usetikzlibrary{decorations.pathmorphing}
\definecolor{darkmagenta}{rgb}{0.55, 0.0, 0.55}
\definecolor{darkblue}{rgb}{0.0, 0.0, 0.55}
\definecolor{darkred}{rgb}{0.7, 0.0, 0.3}

\def\tmu{\tilde{\mu}}
\def\S{{\cal{S}}}

\def\1{1\hspace{-4pt}1}
\def\j1{\widetilde{1\hspace{-4pt}1}}
\usepackage{bbm}

\def\V{{\mathtt{V}}}
\def\nC{{\mathtt{C}}}
\def\tq{{\tilde{q}}}

\def\bec{\begin{center}}
\def\ec{\end{center}}

\def\cL{{\cal L}}
\def\cE{{\cal E}}

\def\cC{{\cal C}}

\def\cR{{\cal R}}
\def\cS{{\cal S}}
\def\cQ{{\cal Q}}

\def\lp{\ell_{\rm P}}
\def\ed{\end{document}}

\def\bea{\begin{eqnarray}}
\def\eea{\end{eqnarray}}
\def\ba{\begin{array}}
\def\ea{\end{array}}

\definecolor{rougef}{rgb}{0.7,0,0}
\definecolor{vertf}{rgb}{0,0.6,0}
\definecolor{bleuf}{rgb}{0,0,0.9}

\usepackage{pdfpages}

\usetikzlibrary{decorations.pathmorphing}
\definecolor{lightblue}{rgb}{0.0, 0.44, 1.0}
\definecolor{purple}{rgb}{0.5,0.0,0.5}
\definecolor{wine}{rgb}{0.59, 0.0, 1.0}
\definecolor{darkmagenta}{rgb}{0.56, 0.0, 0.55}
\definecolor{darkblue}{rgb}{0.0, 0.0, 0.55}
\definecolor{darkred}{rgb}{0.7, 0.0, 0.3}
\hypersetup{
	unicode,	
	colorlinks,
	citecolor=lightblue,
	linkcolor=purple,
	urlcolor=lightblue,
	bookmarksopen=true,
	bookmarksnumbered
	}

 \csname
@addtoreset\endcsname{equation}{section}

\begin{document}
\begin{titlepage}
\setcounter{page}{1}
\begin{center}
\hfill
\vskip 0.1cm
{\LARGE\bf Thermal fluctuations of black holes with non-linear electrodynamics and charged Renyi entropy}
\vskip 20pt
{\sc  Gabriel Arenas-Henriquez}\textsuperscript{\tiny\Coffeecup},~
{\sc Felipe Diaz}\textsuperscript{\tiny\Bicycle},~ and
{\sc  Yerko Novoa}\textsuperscript{\tiny\Bat}

{\it \textsuperscript{\tiny\Coffeecup} Centre for Particle Theory,
  Department of Mathematical Sciences, \\[-1mm] Durham University, South Road, Durham, DH1 3LE, UK
\\[1mm]\textsuperscript{\Bicycle}Departamento de Ciencias F\'isicas, Universidad Andr\'es Bello, \\Sazi\'e 2212, Santiago de Chile
    \\[1mm]\textsuperscript{\tiny\Bat}Departamento de Filosof\'ia, Universidad de Santiago de Chile,\\ Avenida Libertador Bernardo O'Higgins 3677, Santiago, Chile}
\vskip 40pt
{\bf Abstract} \\

We extend the charged Renyi entropy to a more general holographic scenario. Coupling an arbitrary non-linear electrodynamics Lagrangian density to AdS gravity, we analyse the thermodynamic features of non-linearly charged hyperbolic black holes and the thermal fluctuations in the grand canonical ensemble. We provide a general form for the relevant holographic quantities that describes a CFT with a global $U(1)$ symmetry in terms of horizon data and we compute the first thermal fluctuation of the charged Renyi entropy. We demonstrate the validity of the formulae through an analytic example; the Coulomb source in $2+1$ dimensions.  We  propose this model to be dual to charged free bosons in $1+1$ dimensions. The corrections generates a subleading logarithmic divergence in the entanglement entropy which appear in some Condensed Matter systems with spontaneous symmetry breaking due to IR effects in the ground state. We comment on the possibility of interpreting these results in terms of holography beyond the saddle point approximation. 
\end{center}

\vspace*{0.8cm}
\begin{flushleft}
{\textsuperscript{\tiny\Coffeecup}\href{mailto:gabriel.arenas-henriquez@durham.ac.uk}{\texttt{\footnotesize gabriel.arenas-henriquez@durham.ac.uk}}}\\[-2mm]
{\textsuperscript{\tiny\Bicycle}\href{mailto:f.diazmartinez@uandresbello.edu}{\texttt{\footnotesize f.diazmartinez@uandresbello.edu}}}
\\[-2mm]{\textsuperscript{\tiny\Bat}\href{mailto:yerko.novoa@usach.cl}{\texttt{\footnotesize yerko.novoa@usach.cl}}}
\end{flushleft}
\end{titlepage}
{\small\tableofcontents }

\section{Introduction}

The AdS/CFT correspondence \cite{Maldacena:1997re,Witten:1998qj, Gubser:1998bc} maps the gravity partition function in anti-de Sitter (AdS) space to a conformal field theory (CFT) in one dimension lower. The original proposal matches on-shell highly symmetric strongly coupled gauge theories at large $N$ with low energy limits of String theory, implying a UV/IR correspondence \cite{Susskind:1998dq, Peet:1998wn}. After a long and successful path, it has been extended to large class of quantum field theories that are described in terms of IR saddles of quantum gravity in hyperbolic spaces, showing a surprising relation between black hole physics and critical many-body quantum systems, see for instance \cite{McGreevy:2009xe} and reference therein. 

The String theory partition function contains higher-curvature terms in the IR limit such that considering Kaluza--Klein reductions can lead to non-linear electrodynamics (NED) coupling to the Einstein sector \cite{Gibbons:2000xe}. Moreover, non-linear kinetic terms for the vector fields also appear in quantisation of String actions \cite{Fradkin:1985qd}. Therefore, NED Lagrangians, originally introduced as an alternative to Maxwell electrodynamics to solve problems in quantum electrodynamics, can be considered in the duality and allows to extend the possible holographic field theories. For  a nice review on NED and its applications see \cite{Sorokin:2021tge}.

An important matter is to see how the correspondence is extended beyond the classical saddle in the gravity side, such that quantum corrections of the String partition function can modify the properties of the dual CFT. An interesting holographic quantity that allows to explore this issue is the entanglement entropy \cite{Ryu:2006bv, Ryu:2006ef} which has lead to understand properties of spacetime itself \cite{VanRaamsdonk:2010pw}. The quantum corrections to the holographic entanglement entropy have been proposed in \cite{Faulkner:2013ana, Engelhardt:2014gca} where the extra term appearing are of zero order in the gravity coupling.  

In \cite{Belin:2013dva}, an extension of holographic entanglement was introduced when the quantum field theory has a global charge. For a spherical entangling surface, the bulk dual corresponds to charged static topological black holes in the grand canonical ensemble. The charged entanglement has been also studied for the case when the bulk is coupled to Born--Infeld electrodynamics \cite{Dey:2016pei} and higher-derivative Einstein--Maxwell electrodynamics \cite{Cano:2022ord}. 

The aim of this paper is twofold; first we generalise the charge Renyi entropy to the case where the bulk theory contains non-linear sources, and to treat the issue of Gaussian fluctuations in the thermal partition function evaluated on the hyperbolic charged static solution in order to study corrections to the holographic charged Renyi entropy proposed in \cite{Belin:2013dva}. The fluctuations corrects the black hole entropy adding a new subleading term that has been extensively study in the literature, and whose coefficient will depend on the approach to quantum gravity, see for instance \cite{Das:2001ic} and reference therein. This new contribution modifies the IR structure of the Renyi entropy in a very specific way while keeping the UV entanglement unmodified. Interestingly, this behavior has been observed in field theories with spontaneous symmetry breaking (SSB) of a continuous symmetry due to low energy excitations on the ground state \cite{2011PhRvB..84p5134K, Metlitski:2011pr, Alba2013EntanglementSO, Kallin:2014oka, Frrot2016EntanglementEA, Casini:2019kex}. 
This means that, holographically, this contribution cannot be extracted from the classical saddle point of the dual gravitational theory \cite{Jeong:2022zea, Park:2022oek}. 
While the matter fields does not play a relevant role in the particular structure of the corrected Renyi entropy, the consideration of non-linear sources allows us to explore different holographic models beyond the examples of charged AdS black holes known in the literature so far. In particular, we focus in the holographic dual of two-dimensional charged free bosons that we found to match the Renyi entropy and twist operators obtained using Einstein AdS gravity coupled to conformal electrodynamics in the bulk.

The present work is organised as follows:

\paragraph{$\diamond$}In \autoref{section2}, we give a brief review of the standard Renyi entropy and its generalisation to theories with global symmetries. Using the proposal of \cite{Hung:2011nu, Belin:2013dva} for the holographic dual of (charged) Renyi entropy for spherical entangling region, we introduce the main concepts that are going to be used extensively through this manuscript.

\paragraph{$\diamond$} In \autoref{Sec:AdSgeneral}, we extend the gravity dual theory to include non-linearly charged black holes by considering a general NED Lagrangian on top of the Einstein--Hilbert action in asymptotically AdS spacetimes. We analyse, in a general framework, the thermodynamic description and we compute relevant quantities such as the vacuum energy, temperature, and the specific heat. 
Then, in \autoref{SubSec:holog} we obtain the desire holographic data in presence of the NED Lagrangian, such as the conformal weights of the twist operators, the magnetic response, and the charged Renyi entropy.

\paragraph{$\diamond$} In \autoref{Sec:ThermalFluctutations}, we review the method of \cite{Mahapatra:2011si} to compute thermal fluctuations in the gravity partition function in the grand canonical ensemble, and discuss how the new terms modify the Renyi entropy, as proposed in \cite{Mahapatra:2016iok}, and the twist operators conformal weights in general NED.

\paragraph{$\diamond$} In \autoref{Sec:Coulomb}, we present a specific example in three-dimensional gravity in which the NED theory is conformal and the black hole solution is asymptotically AdS. This last feature differs with the static charged BTZ \cite{Martinez:1999qi} which contains a logarithmic term in the metric that breaks the asymptotic symmetries of the solution.
The resulting Renyi entropy and conformal weights reproduce the result of a charged free boson in two-dimensional CFT, which is dual to the Einstein--Maxwell theory in higher dimensions, but not in $d=2$ as can be seen in \cite{Belin:2013dva}.
Furthermore, we compute the thermal fluctuations of system generating the extra logarithmic term in the thermal entropy. This new term generates a subleading logarithmic divergence in the Renyi and entanglement entropy, which resembles the entanglement entropy of SSB systems. Therefore, the coefficient appearing in the logarithmic corrections to the black hole thermal entropy can be interpreted as the number of Goldstone modes in the dual CFT appearing in a SSB of a continuous symmetry. 

We conclude with some discussion of the main results presented in this paper and further directions. 

\section{Holographic charged Renyi Entropy} \label{section2}
Renyi entropy corresponds to a quantity that generalises various notions of entropy while preserving additivity for non-dependant events, and has shown to be an important quantity in several areas of physics. Particularly, Renyi entropy gives information of the entangled structure in many-body quantum ground states. Considering a system subdivided in $A$ and $B$ by the entangling region $\Sigma$, tracing over the states in $B$ defines the reduced density matrix $\rho_A = {\rm Tr}_B \rho$~, where $\rho$ corresponds to the normalised density matrix of the full system. Then, the Renyi entropy of the configuration is
\begin{align}
    S_n = \frac{1}{1-n}\log {\rm Tr}_A \rho_A^n~,
\end{align}
and satisfies the following inequalities \cite{beck_schogl_1993, zyczkowski2003renyi}
\begin{align}\label{ineq}
    \partial_n S_n \leq 0~,\qquad \partial_n \left(\frac{n-1}{n}S_n\right) \geq 0~,\qquad \partial_n ((n-1)S_n) \geq 0~,\qquad \partial_n^2 ((n-1)S_n)\leq 0~. \end{align}
Different limits of the Renyi entropy give useful information about the reduced density matrix; the $n\to1$ limit gives the Von Neumann entropy 
\begin{align}\label{RenyiEntropy}
    S_{\rm E} := \lim_{n\to1}S_n = -{\rm Tr}_A \rho_A \log \rho_A~,
\end{align}
that defines the entropy of entanglement between the degrees of freedom in both regions. Other two interesting limits are
\begin{align}
    S_0 := \lim_{n\to 0}S_n = \log{\mathbf{d}}~,\qquad S_\infty := \lim_{n\to \infty}S_n = -\log{\lambda_{\rm max}}~,
\end{align}
where $\mathbf{d}$ is the number of non-zero eigenvalues of $\rho_A$, that is, the dimension of the subspace of the Hilbert space in the region $A$ that participates in the entanglement, and $\lambda_{\rm max}$ is the largest eigenvalue of $\rho_A$. An interesting fact is that the second Renyi entropy, $S_2 := \lim_{n\to2}S_n$ has been measured experimentally in \cite{Islam:2015mom} corresponding to the bipartition entanglement between two identical systems, and the proposal to extend this experiment for the inclusion of global symmetries can be found in \cite{Goldstein:2017bua}.

Furthermore, the Renyi entropy has a thermal interpretation by considering a field theory at temperature $T_{0}$ and with free energy functional $F(T_{0})$. For a Gibbs state at temperature $T=T_{0}/n$, the Renyi entropy can be defined as \cite{Baez:2011upp}
\begin{align}
    S_n = \left(1-\frac{1}{n}\right)^{-1}\frac{1}{T_0}\left[F(T_{0}) - F(T_0/n)\right]~,
\end{align}
and can easily be related to the thermal entropy of the system by means of standard thermodynamic relations, viz.
\begin{align}\label{RenyiTh}
    S_n = \left(1-\frac{1}{n}\right)^{-1}\frac{1}{T_0}\int^{T_0}_{T_0/n} S_{\rm thermal}(T) dT,
\end{align}
such that the Renyi parameter is associated with the decreasing of the temperature.

In practice, obtaining the Renyi entropy is rather complicated but, for certain field theories, it can be done through the Replica trick \cite{Calabrese:2004eu}. This allows to replace the computation of the $n$-th power of the reduced density matrix, $\rho_A^n$~, with obtaining the partition function of the original field theory on the replicated manifold, $Z_n$~, which contains $n$-copies of the original manifold glued cyclically along the $n$-fold branch-cuts introduced in the entangling region, i.e.,
\begin{align}
    {\rm Tr}_A \rho_A^n =  \frac{\int \left[\prod_{i =1}^n{\cal D
    }\phi_i\right] \prod_j \sigma_{j,n}(\omega_k){\rm exp}\{-\sum_{k = 1}^n I_k[\phi]\}}{\int \left[\prod_{i =1}^n{\cal D
    }\phi_i\right] ~{\rm exp}\{-\sum_{k=1}^nI_k[\phi]\}} = \frac{Z_n}{Z_1^n}~,
\end{align}
with $\sigma_n$, usually referred as twist operators, located at the end-points $\omega_k$ of the $n$-fold branch-cuts \cite{Calabrese:2004eu, Hung:2011nu, Hung:2014npa}. Then, the Renyi entropy \eqref{RenyiEntropy} becomes
\begin{align}
    S_n = \frac{1}{1-n}(n\log Z_1 - \log Z_n)~.
\end{align}

The computation of the Euclidean path integral on a complicated Riemann surface also correspond to compute correlation function of twist operators, which is indeed something achievable in quantum field theories. For a two-dimensional CFT, the twist operators are primaries with scaling dimension
\begin{align}
    h_n = \frac{c}{12}\left(n-\frac{1}{n}\right)~,
\end{align}
where $c$ is the central charge of the theory, and in higher-dimensional CFT a power expansion of the conformal weights $h_n$ in the Renyi index $n$ around $n=1$ found to be proportional to the structure coefficients of the theory \cite{Hung:2011nu, Perlmutter:2013gua, Belin:2013uta, Hung:2014npa}.

Holographically, Ryu and Takayanagi \cite{Ryu:2006bv, Ryu:2006ef} proposed, and later proved by Lewkowycz and Maldacena \cite{Lewkowycz:2013nqa}, that the entanglement entropy of a field theory is dual to extremising the area of a co-dimension 2 surface whose boundary corresponds to be conformal to the entangling region $\Sigma$. This idea generalises the Bekenstein--Hawking area law for black hole entropy in the case in which the minimal surface equals the black hole horizon.

For the case of a zero-temperature $d$-dimensional CFT in flat space with spherical entangling region,  Casini, Huerta, and Myers \cite{Casini:2011kv} showed that the density matrix can be mapped to a thermal density matrix on a CFT in a hyperbolic cylinder $\mathbb{R}\times {\mathbb H}^{d-1}$ of radius $R$, which corresponds to a slicing of AdS space, at temperature\footnote{See \cite{Jensen:2013lxa} for the extension of the CHM map to CFTs with planar boundaries or planar defects.} 
\begin{align}
    T_0 = \frac{1}{2\pi R}~,
\end{align}
such that the entanglement entropy of the non-thermal CFT is mapped to the thermal entropy of the CFT on the hyperbolic cylinder at temperature $T_0$.
This is usually known as the CHM map, and corresponds to a unitary transformation of the density matrix such that, after taking traces, the quantum information of the system remains invariant. From the bulk point of view, one can use the CHM map to relate the entanglement entropy of a non-thermal CFT with spherical entangling region to the Bekenstein--Hawking entropy of a static topological black hole with Hawking temperature $T_0$. Therefore, this map in the gravity side corresponds to a coordinate transformation which renders the black hole horizon into the Ryu--Takayangi minimal surface. 

The holographic Renyi entropies are usually mapped to gravity duals on non-smooth geometries \cite{Dong:2016fnf} due to the Replica trick. These geometries contain conical singularities \cite{Fursaev:1995ef} that must be included in the gravity partition function. The CHM map reduces the computation to expressing the formula \eqref{RenyiTh} in terms of the black hole entropy and temperature defined by the horizon radius $r_h$. Therefore, the equation \eqref{RenyiTh} from the point of view of the bulk,  becomes
\begin{align}\label{RenyiST}
    S_n ={}& \left(1-\frac1n\right)^{-1}\frac{1}{T_0}\int^{x_1}_{x_n}S(x)\partial_x T(x) dx~, \\ ={}& \left(1-\frac1n\right)^{-1}\frac{1}{T_0}\left[S(x)T(x)\rvert_{x_n}^{x_1} - \int_{x_n}^{x_1}T(x)\partial_x S(x) dx \right],
\end{align}
where $x := r_h/L$~,  with $L$ the AdS radius, $S(x)$ and $T(x)$ are the black hole entropy and temperature respectively, and $x_n$ represents the change of the horizon radius in terms of the change of temperature $T(x_n) = T_0/n$~. 

One can be a bit more curious and wonder what happens with the above definitions in a CFT with a continuous global symmetry. This question has been already answered in \cite{Belin:2013uta} for the case of a Wilson line that encircles the entangling surface. The conserved global charge adds a fixed chemical potential $\mu$ to the free energy without modifying the form of \eqref{RenyiST} and the twist operators get a dependence on the chemical potential due to the magnetic flux carried by the entangling surface. Their conformal dimension $h_n(\mu)$ is extracted from the leading order behaviour of the correlators 
\begin{align}
    \langle T_{ab} \sigma_n(\mu)\rangle = -\frac{h_n}{2\pi}\frac{\delta_{ab}}{y^d}~,\qquad \langle T_{ij}\sigma_n(\mu)\rangle = -\frac{h_n}{2\pi} \frac{d n_i n_j - (d-1)\delta_{ij}}{y^2}~,
\end{align}
where $n_i$ are unit vectors orthogonal to the insertion of the twist operators, $y$ is the perpendicular distance between the insertion of the stress tensor and the twist operator, and the indices $(a,b)$ denote tangential directions to the twist operators, and indices $(i,j)$ denote normal directions.  In fact, the conformal weights $h_n(\mu)$ can be computed in terms of the energy density ${\cal{E}}(T,\mu)$ of the CFT \cite{Belin:2013uta} as
\begin{align}
    h_n(\mu) =  \frac{2\pi n}{d-1}R^d \left({\cal{E}}(T_0,0)-{\cal{E}}(T_0/n,\mu)\right)~. \label{henergy}
\end{align}
In the bulk, the difference of energy densities is extracted from the $(t,t)$-component holographic stress tensor \cite{Balasubramanian:1999re} which in the case of the topological black hole leads to
\begin{align}
    h_n(\mu) = -\pi n \left(\frac{L}{\lp}\right)^{d-1}m(x_n,\mu)~,
\end{align}
where $m$ is a parameter related with the mass of the black hole written in terms of the chemical potential and $x_{n}$. The last expression holds independently of the gauge sector of the bulk theory. The gauge field $A_{\mu}$ in the bulk is dual to a current $J_{i}$ at the boundary from which the leading order divergence of the correlator
\begin{align}
    \langle J_i \sigma_n(\mu)\rangle  = - \frac{i k_n}{2\pi}\frac{\epsilon_{ij}n^j}{y^{d-1}}~,
\end{align}
where $\epsilon_{ij}$ is the volume form of the space transverse to the entangling region, and the structure coefficient $k_n(\mu)$ is referred as the magnetic response, given by  
\begin{equation}\label{magneticresponse}
k_n(\mu) = 2\pi n R^d \rho(x_n,\mu)~,
\end{equation}
where $\rho(n,\mu)$ corresponds to the charge density of the theory appearing in the first law of thermodynamics
\begin{align}
    d\cE = TdS + \frac{\mu}{2\pi R} d\rho  \Rightarrow \frac{\mu}{2\pi R} = \left(\frac{\partial\cE}{\partial\rho}\right)_S~.
\end{align}
Additionally, the coefficient $C_T$ in the leading singularity of the stress tensor two-point function, which encodes information about the central charges of the theory, can be expressed as \cite{Hung:2011nu, Perlmutter:2013gua},
\begin{align}
    C_{T} = \frac{d-1}{2\pi^{\frac{d-2}{2}}\Gamma(d/2)} h_{1,0}~,
\end{align}
where $\Gamma(d)$ is the Gamma function and $h_{1,0}$ is a coefficient in an expansion of the conformal weight around $n=1$ and $\mu= 0$, 
\begin{align} \label{expansionh}
    h_n(\mu) = \sum_{a,b}\frac{1}{a!b!}h_{a,b}(n-1)^a \mu^b~,\qquad h_{a,b} = \partial_n^a \partial_\mu^b h_n(\mu)\Big\rvert_{{\substack{n=1\\\mu=0}}}~.
\end{align}
Thus, the coefficients of the most singular order of the correlators, which encodes most of the crucial information about the CFT, can be extracted by expansions of the conformal weights of twist operators. In the next section we will describe the bulk theory and the holographic setup. 

\section{AdS Gravity and non-linear Electrodynamics} \label{Sec:AdSgeneral}
For the bulk theory, we consider the Einstein--Hilbert AdS action coupled to an arbitrary NED density $\cL(\cS)$ that depends on the Lorentz invariant scalar $\S = F_{\mu\nu}F^{\mu\nu}$~, where $F_{\mu\nu}$ is the field strength of the $U(1)$ gauge field $A_\mu$~,
\begin{align}
    I = \int d^{d+1}x\sqrt{g}\left(\frac{1}{2\lp^{d-1}}\left(\cR + \frac{d(d-1)}{L^2}\right) + \cL(\S)\right)~, 
\end{align}
where $\cR = \cR^{\mu\nu}g_{\mu\nu}$ is the Ricci scalar, and $L$ is the characteristic AdS radius. The equations of motion reads 
\begin{align}
    E^\mu_\nu\equiv {}& \cR^\mu_\nu - \frac12\delta^\mu_\nu (\cR - 2\Lambda) - \lp^{d-1}T^\mu_\nu = 0~, \\ \label{Maxeq} E^\mu \equiv{}& \nabla_{\nu}\left(F^{\mu\nu}\frac{\partial \cL}{\partial \cS} \right) = 0~,
\end{align}
where the NED energy-momentum tensor is
\begin{align}\label{Tmunu}
    T^\mu_\nu = \delta^\mu_\nu \cL - 4\frac{\partial \cL}{\partial \cS}F^{\mu\alpha}F_{\nu\alpha}~.
\end{align}
Solving the field equations for a static ansatz\footnote{The normalisation chosen is such that  for $k=-1$ \cite{Hung:2011nu} the boundary line element becomes conformally equivalent to $ds^2=-dt^2 + R^2dH_{d-1}^2$ corresponding to a hyperbolic cylinder of radius $R$ where $dH^2_{d-1}$ is the volume element of the $(d-1)$ Lobachevsky disc.}
\begin{align}
    ds^2 = -f(r)\frac{L^2}{R^2}dt^2 + \frac{dr^2}{f(r)} + r^2\gamma_{mn}dy^m dy^n~,
\end{align}
where the curvature of the horizon is labeled by the topological parameter $k = 0,1,-1$ for flat, spherical, or hyperbolic case, respectively. The gauge field ansatz
\begin{align}
    A_\mu = \left(\frac{L}{R}\phi(r)-\frac{\mu}{2\pi R}\right)\delta^t_\mu~,\qquad \mu \equiv 2\pi L\phi(r_h)~,
\end{align}
where $\mu$ is the chemical potential chosen such that the gauge potential vanishes at the horizon, generates an electric field $E(r) = -\partial_r \phi(r)$~ and the associated field strength
\begin{align}
    F_{\mu\nu} = \frac{L}{R}E(r)\left(\delta^t_\mu\delta^r_\nu - \delta^t_\nu\delta^r_\mu\right)~,
\end{align}
such that
\begin{align}
    \S = -2E^2~.
\end{align}
We can plug the metric function $f(r)$ into the equations of motion and using the fact that $T^t_t = T^r_r$, we get
\begin{align}
    E^t_t = E^r_r = \frac{d-1}{2r^2}\left(rf'(r) + (d-2)(f(r)-k) - d\frac{r^2}{L^2}\right) - \lp^{d-1}T^r_r = 0,
\end{align}
that is solved for
\begin{align}\label{fgen}
    f(r) = \frac{r^2}{L^2} + k - \frac{m}{r^{d-2}} +\frac{2\lp^{d-1}}{d-1}\frac{\cQ}{r^{d-2}}~,
\end{align}
where
\begin{align}\label{cQ}
    \cQ(r) := \int_\infty^r du~u^{d-1} T^r_r = \int_\infty^r du\left(u^{d-1}\cL(\cS) - 4\tq E\right)~.
\end{align}
The parameter $m$ is an integration constant that can be related to the black hole mass through the Noether-Wald formalism. The electric charge of the black hole can be obtained as the Noether charge associated to the $U(1)$ gauge symmetry \cite{Miskovic:2010ui}. The black hole horizon is given by the root of $f(r_{h})=0$. We can use this to write the mass as a function of the horizon,
\begin{align}\label{mass}
    m =  \frac{r_h^{d-2}}{L^2}\left(r_h^2 + kL^2\right) + \left(\frac{2\lp^{d-1}}{d-1}\right)\frac{\cQ(r_h)}{r_h^{d-2}}
\end{align}
and thus, express the metric function as
\begin{align}
    f(r) = \frac{r^2}{L^2} + k + \frac{2\lp^{d-1}}{d-1}\frac{1}{r^{d-2}}\cQ(r) - \left(\frac{r_h}{r}\right)^{d-2}\left(\frac{r^2_h}{L^2} + k +\frac{2\lp^{d-1}}{d-1}\frac{1}{r^{d-2}_h}\cQ(r_h) \right)~.
\end{align}
The black hole temperature is
\begin{align}
    T = \frac{T_0}{2}Lf'(r_h) = \frac{T_0}{2}\frac{L}{r}\left[d\frac{r_h^2}{L^2}+(d-2)k  +  \frac{2\lp^{d-1}}{d-1} r_h^2 T^r_r({r_h}) \right]~,
\end{align}
such that extremality occurs at
\begin{align}
T_r^r(r_h) = -\frac{d-1}{2\lp^{d-1}r_h^2}\left(d\frac{r^2_h}{L^2} +k(d-2)\right)~,    
\end{align}
and the Wald entropy corresponds to 
\begin{align} \label{waldentropy}
    S^{(0)} = \V \left(\frac{r_h}{L}\right)^{d-1}~,\qquad \V \equiv {2\pi}\left(\frac{L}{\lp}\right)^{d-1}V_\Sigma~,
\end{align}
where $V_\Sigma$ is the volume of the horizon that for the hyperbolic case ($k=-1)$
\begin{align}
V_{\mathbb{H}^{d-1}} = \int_{\mathbb{H}^{d-1}} d\Sigma_{d-1} \sim \frac{\Omega_{d-2}}{d-2}\frac{R^{d-2}}{\delta^{d-2}}+\dots~,\qquad \Omega_{d-2} \equiv \frac{\pi^{\frac{d-2}{2}}}{\Gamma(d/2)}~,
\end{align}
is divergent, regularised by inserting a short-distance cutoff $\delta$, such that the Renyi entropy universal divergences are already encoded in this quantity  \cite{Casini:2011kv}. For $d=2$~, the volume logarithmically diverges
\begin{align}
    V_{\mathbb{H}^1} = 2\log\left(\frac{2R}{\delta}\right)~,
\end{align}
which encapture the logarithmic divergences in two-dimensional CFT entanglement entropy \cite{Calabrese:2004eu}.
We can integrate the gauge field equation \eqref{Maxeq} to obtain a generalised Gauss law
\begin{align}\label{Maxt}
    E^t = 0 = \partial_r \left( r^{d-1}E(r)\frac{\partial \cL}{\partial \cS} \right)\Rightarrow E(r)\frac{\partial \cL}{\partial \cS}\Bigg\rvert_{\cS=-2E^2} = -\frac{\tilde{q}}{{r^{d-1}}}~,
\end{align}
where $\tq$ can be related with the physical electric charge of the solution.

It is known that for black holes with a non-linear electrodynamic source, the Smarr relation is not satified,  see \cite{Balart:2017dzt} for a discussion. Nonetheless, in \cite{Miskovic:2010ey} was noticed that instead of relying on the Smarr relation to derive the first law of black hole thermodynamics, it is possible to get the quantum statistical relation from the regularised Euclidean action
\begin{align}
    G \equiv T I = {\cal E} - TS - \frac{\mu}{2\pi R}\rho~.
\end{align}
It follows that the first law of black hole thermodynamics is
\begin{align}
    d{\cal E} = TdS + \frac{\mu}{2\pi R}d\rho~,
\end{align}
where the total energy of the system ${\cal E} = M + \frac{1}{2}(1+(-1)^{d})E_{\rm vac}$~, with $M$ as the ADM mass, gets a contribution from the Casimir energy of the dual field theory when $d$ is even \cite{Horowitz:1998ha},
\begin{align}
   E_{\rm vac} = -\frac{2V_\Sigma}{\sqrt{\pi}}\frac{\Gamma((d+1)/2)}{\Gamma(d/2)}\frac{(-L)^{\frac{d}{2}-1}}{\lp^{d-1}}  \lim_{r\to\infty} \int_{0}^1 dy\left(f - \frac{r}{2}f'\right)\left(y^2\left(\frac{r^2}{L^2} - f\right)+k \right)^{\frac{d}{2}-1}~,
\end{align}
that corresponds to the vacuum energy of global AdS spaces in odd dimensions. Assuming finiteness of the total energy, the Casimir energy is
\begin{align}
    E_{\rm vac} = (-k)^{\frac{d}{2}}\frac{L^{\frac{d}{2}-1}}{\lp^{d-1}}V_\Sigma = \frac{(\pi L)^{\frac{d}{2}-1}}{(d-2)\lp^{d-1}\Gamma(d/2)}\frac{R^{d-2}}{\delta^{d-2}}+\dots~,
\end{align}
where the second equality corresponds to the hyperbolic case $k=-1$. We will use the generalised quantum statistical relation to see how the thermal fluctuations modify the known holographic quantities such as the conformal weight of the twist operators. 

The black hole specific heat 
\begin{align} \label{especific}
    {\cal{C}} = T \left(\frac{\partial}{\partial T}S^{(0)}\right) = T(d-1)\V x^{d-2}\frac{\partial x}{\partial T}~,
\end{align}
where
\begin{align} \label{dxdt}
    \frac{\partial x}{\partial T} = \frac{2}{T_0}\left[d -\frac{k(d-2)}{x^2} + \frac{2\lp^{d-1}L^2}{d-1}\left(T^r_r + x\partial_x T^r_r  \right) \right]^{-1}, 
\end{align}
is expressed in terms of the NED energy-momentum tensor evaluated at the horizon $x$. In other words, the specific heat is
\begin{align}\label{specificheat}
    {\cal{C}} ={}& \frac{2(d-1)\V x^{d}T}{T_0\left[dx^2 - k(d-2) + \frac{2\lp^{d-1}L^2}{d-1}x^2\left(T^r_r + x\partial_x T^r_r  \right)\right]}~,
\end{align}
that by using the entropy formula \eqref{waldentropy} is cast in the same footing as it is presented in the rest of the literature for particular NED theories (see for instance \cite{Gonzalez:2009nn, Cai:2004eh}), 
\begin{align}
    {\cal C} = (d-1)S^{(0)}\left[\frac{dx^2 + (d-2)k + \frac{2L^2\lp^{d-1}}{d-1}x^2 T^r_r}{dx^2 - (d-2)k + \frac{2L^2\lp^{d-1}}{d-1}x^2\left(T^r_r + x\partial_x T_r^r\right)}\right]~.
\end{align}
Requiring positive heat capacity implies that,
\begin{align}
    T^r_r(x) + x\partial_x T^r_r(x) \geq - \frac{d-1}{2\lp^{d-1} L^2}\left(d - \frac{k(d-2)}{x^2} \right)~, \label{localstabilitycondition}
\end{align}
where all terms are evaluated at $\cS = -2E^2$~.
Additionally, the weak energy condition $\rho = -T_{\mu\nu}u^\mu u^\nu \geq 0$ puts an extra constraint in the NED energy-momentum tensor,
\begin{align}
    T^r_r = T^t_t = \cL + 4\frac{\partial \cL}{\partial \cS}E^2 \geq 0~.
\end{align}
Then
\begin{align}\label{conditionTrr}
     \partial_x T^r_r(x) \geq - \frac{d-1}{2\lp^{d-1}L^2 x}\left(d - \frac{k(d-2)}{x^2} + \frac{2\lp^{d-1}L^2}{d-1}T^r_r\right)~,
\end{align} 
which for the topological black hole with hyperbolic horizon $k = -1$, the r.h.s. of the inequality is always negative. 
As we saw in section \ref{section2}, the charged Renyi entropy for a CFT is defined in the grand canonical ensemble, i.e., the chemical potential is fixed. Therefore, in order to obtain the logarithmic corrections of the charged Renyi entropy, we are going to analyse thermal fluctuations in the grand canonical ensemble where the stability condition is modified \cite{Chamblin:1999hg}, and the value of the heat capacity will play an important role when particularising the result to different black hole solutions. 

\subsection{Holographic Quantities}\label{SubSec:holog}
Having defined the relevant bulk expressions, we proceed with the dual boundary description. We follow to list the relevant holographic quantities, which in fact, the NED contributions  to the boundary data can be expressed in terms of the $rr$-component of the  energy-momentum tensor evaluated at the horizon.

The conformal weights of the twist operators
\begin{align} \label{conformalweight}
    h_n(\mu) = \pi n \left(\frac{L}{\lp}\right)^{d-1}\left(x_n^{d-2}L^{d-2}(1-x^2_n) - \left(\frac{2\lp^{d-1}}{d-1}\right)\frac{\cQ(x_n,\mu)}{\left(L x_n\right)^{d-2}}\right)~,
\end{align}
and the magnetic response \eqref{magneticresponse} can be obtained with the generalised charge density  \cite{Miskovic:2010ey} 
\begin{align} \label{generalisedchargedensity}
    \rho(x_n,\mu) = -4\left(\frac{L}{R}\right)^{d-2}x_n^{d-1}\left(E \frac{\partial\cL}{\partial\cS}\right)\Bigg\rvert_{\cS=-2E^2,~ x = x_n}~,
\end{align}
and the central charge of the dual theory can be obtained by taking a series expansion of the structure coefficients around $n=1$~.
All the holographic quantities depend on $x_n$ which corresponds to the largest solution to $T(x_n,\mu) = T_0/n$, viz. 
\begin{align} \label{xngeneral}
   nx_n^2\left( d + \frac{2L^2\lp^{d-1}}{(d-1)}T^r_r(x_n,\mu)\right) - 2x_n + kn(d-2) = 0~,
\end{align}
with $k=-1$. The last equation cannot be always solved with analytic methods for the horizon radius $x_{n}$. As a matter of fact, for most of the lagrangians $\cL(\S)$, \eqref{xngeneral} is solved numerically. The same occurs for the chemical potential; for each NED the chemical potential can be obtained numerically in terms of the electric charge and then be replaced in the desire quantity\footnote{See for instance \cite{Dey:2016pei} for Born--Infeld electrodynamics}. 

Finally, the holographic charged Renyi entropy is obtained by evaluating \eqref{RenyiST} leading to
\begin{align}\label{Sngeneral}
    S_n ={}& \frac{n}{n-1}\frac{\V}{2}\left\{x^{d-2} + x^d + \frac{2\lp^{d-1}L^2}{d-1}\left(\cQ(x,\mu) + \int x^{d}\partial_x T^r_r(x,\mu)dx \right)\right\}\Bigg\rvert_{x_n}^{x_1} ~,
\end{align}
and the Renyi inequalities \eqref{ineq} can be checked using
\begin{align}
    \partial_n S_n = -\frac{\partial x_n}{\partial n}\left[(d-2)x^{d-3}_n + dx^{d-1}_n + \frac{2\lp^{d-1}L^2}{d-1}x_n^{d-1}\left(T^r_r(x_n) +x_n\partial_{x_n} T^r_r(x_n)\right)\right]~,
\end{align}
where
\begin{align}
    \frac{\partial x_n}{\partial n} = \frac{x^2_n\left(d + \frac{2\lp^{d-1}}{d-1}T^r_r(x_n) \right) -(d-2)}{2 - 4nx_n\left[d +\frac{2\lp^{d-1}}{d-1}\left(T^r_r(x_n) + \frac{x_n}{2}\partial_{x_n} T^r_r(x_n)\right) \right]}~.
\end{align}
Finally, the Renyi index can be related with the black hole parameters as
\begin{align}
    n = \frac{2L}{r_h}\left[\frac{dr_h^2}{L^2}-(d-2)+\frac{2\lp^{d-1}}{d-1}r^2_h T_r^r(r_h) \right]^{-1}~,
\end{align}
which can be used to check phase transitions in the dual theory \cite{Belin:2013dva, Fang:2016ehk}.

In the following section, we will include thermal fluctuations to black hole entropy and use the holographic dictionary to interpret the correction from the CFT perspective.  

\section{Thermal fluctuations and stability}\label{Sec:ThermalFluctutations}
Thermal fluctuations induces logarithmic corrections in the black hole entropy. In the grand canonical ensemble with $K$ fixed chemical potentials $\mu_{i}$, with $i = 1,\dots,K$~, the standard procedure to obtain the corrections requires to consider the partition function as a discrete sum over energy states and take the continuum limit \cite{Gour:2003jj}. The entropy acquires a correction \cite{Mahapatra:2011si} of the type
\begin{align}
    S = S^{(0)} + \frac12 \log {\rm det}~D_{AB} + \log J~, \label{loggrand}
\end{align}
where $J$ is a Jacobian \cite{Gour:2003jj} that appears in due to the energy spectrum transformation. To compute the new contributions, one assumes that the charges and energy are linear in their respective quantum numbers, and that the symmetric matrix $D_{AB}$ is given by
\begin{align}
    D_{AB} = \frac{\partial}{\partial \chi^{A}}\frac{\partial}{\partial \chi^{B}} \log Z~,
\end{align}
where $\chi^A = \{-\beta\mu_i, \beta\}$ with $\beta = T^{-1}$, are the thermodynamic quantities considered in the fluctuations of the thermal partition function $Z(\beta,\mu_i/T)$. The logarithmic corrections only makes sense if ${\rm det}D_{AB} \geq 0$. This imposes conditions on the thermodynamic quantities such as the heat capacity or Maxwell relations \cite{Ghosh:2021uxg}. For instance, considering only one chemical potential \cite{Chamblin:1999hg}, the determinant is
\begin{align}
    {\rm det} D_{AB} = T^3 {\cal{C}} \kappa - T^3 \mu - T^4\alpha^2~, 
\end{align}
where
\begin{align}
    {\cal C} = T\left(\frac{\partial S}{\partial T}\right)_\mu~,\qquad \kappa = -\frac{1}{T}\left(\frac{\partial N}{\partial \lambda}\right)_T~,\qquad \alpha = \left(\frac{\partial N}{\partial T}\right)_{\mu}~,
\end{align}
with $\lambda = -\mu/T$ and $N$ corresponding to the thermodynamic charge conjugated to the chemical potential (electric charge in our case). Then, the condition on the determinant leads to 
\begin{align}\label{Ccondition}
    {\cal{C}} \geq T\alpha\left(\frac{\alpha}{\kappa} + \frac{\mu}{T} \right)~.
\end{align}
This expression defines thermal local stability of the system. In the canonical ensemble, this condition simplifies to
\begin{align}\label{LogS}
    S = S^{(0)} + \frac12\log\left({\cal{C}} T^2\right)~,
\end{align}
in agreement with \cite{Das:2001ic}.

Consideration of thermal or quantum fluctuations have shown to modify the entropy as 
\begin{align}
    S = S^{(0)} - \nC\log S^{(0)}+\dots~,
\end{align}
and the value of the $\nC$ coefficient depends on the particular black hole solution and the ensemble under consideration, but changes depending on the approach to quantum gravity and allows to test holographic theories beyond the saddle approximation, see for instance \cite{Das:2001ic, Carlip:2000nv, Mukherji:2002de, Gour:2003jj, Banerjee:2010qc, Sen:2012cj, Sen:2012dw,Pourhassan:2015cga, Pathak:2016vfc, Pourhassan:2017kmm, Ghosh:2021uxg}.
We will see how the $\nC$ coefficient can be interpreted in an holographic context by considering thermal fluctuations on NED solutions and check how affect the holographic quantities computed in \autoref{SubSec:holog}.

The thermal fluctuation back reacts into the Renyi entropy \cite{Mahapatra:2016iok, Arenas-Henriquez:2022pyh} and the charged Renyi entropy that becomes
 \begin{align}\label{RenyiCorr}
     S_n ={}& \frac{n}{n-1}\frac{1}{T_0}\int^{x_1}_{x_n} \left(S^{(0)} - \nC\log\left(\V x^{d-1}\right)\right)\partial_x T(x,\mu) dx  \\ \nonumber
      ={}& \frac{n}{n-1}\left\{\frac{\V}{2}\left( x^{d-2}  + x^d\right) - \frac{\nC}{2}\left[\frac{(2-d(1-x^2))(d-1) + (2-d(1+x^2))\log\V x^{d-1} }{x}\right]\right\}\Bigg\rvert_{x_n}^{x_{1}} \\ \nonumber +{}& \frac{n}{n-1}\left(\frac{\lp^{d-1}L^2}{d-1}\right)\int^{x_1}_{x_n}\left(\V x^{d-1} - \nC\log\V x^{d-1}\right)\left(T^r_r(x,\mu) + x\partial_x T^r_r(x,\mu)\right)dx~,
 \end{align}
where all the quantities are evaluated at fixed chemical potential prior to integration. The first line of \eqref{RenyiCorr} corresponds to the universal term coming from the Einstein--Hilbert action, while the second line contains the contribution of the matter sector. Notice that turning off the gauge field and taking $\nC= 3/2$,  \eqref{RenyiCorr} reduces to the result obtained in \cite{Mahapatra:2016iok}.  It is also important to notice that the last expression has a smooth $\mu \to 0$ limit. In fact, we can see from the first line that there is no explicit dependence on the chemical potential, being relevant just for the value of $x$. This means that the structure of the next to leading order contribution in the Renyi entropy is fixed by the gravity solution, and not by the matter source. Later on, we will examine this in detail for a concrete example where we will see that this structure is present in systems where there has been some spontaneous symmetry breaking of a spatial continuous symmetry.

The Renyi inequalities \eqref{ineq} are satisfied if the Wald entropy is positive \cite{Pastras:2014oka}
as can be seen from 
\begin{align}
    \partial_n\left(\frac{n-1}{n}S_n(\mu)\right) = \frac{1}{n^2}S(T_0/n,\mu)~ \geq 0~.
\end{align}
Therefore, the Renyi inequalities can be violated if $\nC\geq \frac{S^{(0)}}{\log S^{(0)}}$. In the grand canonical ensemble $\nC$ 
can take positive or negative values, but in the canonical ensemble $\nC$ is indeed positive \cite{Das:2001ic}, opening the possibility for negative entropy. Nonetheless, as the value of $\nC$ is usually small compared with the value of the entropy, the bound can be violated only for microscopic black holes \cite{Scardigli:1999jh}. One can think of this coming from an evaporation process of a black hole where at late times becomes so small that the thermal fluctuations start dominating over the classical saddle causing a ``negative'' entropy. However, this description is not accurate in such scenario. In fact, these microscopic black holes appear to be thermally unstable \cite{Jizba:2009qf}, which prevents us from describing the thermal fluctuations with this method. A more precise description would require to consider how the geometry itself reacts due to quantum effects modifying the thermodynamic quantities in the regime where the solution is locally stable \cite{Pourhassan:2022opb}. Therefore, the Renyi inequalities of the dual theory can be understood as the thermal stability condition of the black hole \cite{Nakaguchi:2016zqi} even when logarithmic corrections are considered.

Let us consider, for a moment, only the gravity sector, i.e. $\mu = 0$~. We can compute the conformal weight of the twist operator by integrating the first law \cite{Hung:2011nu, Dey:2016pei}. This leads to
\begin{align}
    h_n(\mu=0) ={}& \frac{n}{(d-1)V_{\mathbb{H}^{d-1}}}\int^{x_1}_{x_n} T(x)\partial_x S(x) dx \nonumber \\ ={}& \pi n \left(\frac{L}{\lp}\right)^{d-1}\left( x_n^{d-2} - x_n^{d} \right) - \frac{n\nC}{2V_{\mathbb{H}^{d-1}}}\frac{(1 - x_n)(d x_n + d-2)}{x_n}~, 
\end{align}
where
\begin{align}
    x_n = \frac{1}{dn}\left(1 + \sqrt{1-2dn^2 + d^2n^2}\right)~.
\end{align}
Thus, 
\begin{align}
    \partial_n h_n\rvert_{n=1} = \frac{1}{d-1}\left(2\pi\left(\frac{L}{\lp}\right)^{d-1} - \frac{\nC}{V_{\mathbb{H}^{d-1}}}\right)~.
\end{align}
The second term is identified as next to the leading order contribution in the saddle point approximation of the gravity dual \cite{Susskind:1998dq}. From the quantum field theory perspective, it corresponds to an IR correction of the central charge, which is in agreement with the $c$- and $F$-theorem in $d=2$ and $d=4$, respectively \cite{Zamolodchikov:1986gt, Cardy:1988cwa}. The latter is true if and only if, the coefficient $\nC$ is positive, which is indeed required for the black hole to satisfy the generalised second law \cite{Das:2001ic}. 

Extending the result to nontrivial chemical potential we get
\begin{align}
    h_n(\mu) ={}& \frac{n}{(d-1)V_{\mathbb{H}^{d-1}}}\left(\int_{x_n}^{x_1} T(x,\mu)\partial_x S(x,\mu) dx - \frac{V_{\mathbb{H}^{d-1}} \mu}{2\pi R} \rho(x_n,\mu)\right) \\ ={}& \pi n \left(\frac{L}{\lp}\right)^{d-1}\left( x_n^{d-2} - x_n^{d} \right) - \frac{n\nC}{2V_{\mathbb{H}^{d-1}}}\frac{(1 - x_n)(d x_n + d-2)}{x_n}- \frac{n\mu}{2\pi R} \frac{\rho(x_n,\mu)}{d-1}~,
\end{align}
where $x_n = x_n(\mu)$~. In the grand canonical ensemble, the logarithmic correction can have a negative $\nC$ coefficient, such that the entropy is always positive \cite{Gour:1999ta, Major:2001ue, Mahapatra:2011si}.  

In the next section, as an analytic example, we particularise the results obtained for the charged Renyi entropy and its logarithmic corrections for three-dimensional AdS gravity coupled to Coulomb source. We will show how the bulk seems to describe charged free bosons in $1+1$ dimensions, and how the extra subleading logarithmic term in the entanglement resembles the entanglement entropy of systems with SSB. For additional examples, see the Appendix \ref{App:Power Law}.

\section{Holographic charged free boson in 1+1}\label{Sec:Coulomb}
In the usual three-dimensional AdS gravity with Maxwell electrodynamics, the BTZ black hole \cite{Martinez:1999qi} is charged by
\begin{align}\label{btzlog}
   \cQ(r) = - \frac{q^2}{4\lp}\log\left(\frac{r}{\sqrt{mL^2}}\right)~.
\end{align}
Clearly, this solution does not asymptotes to global AdS. From the dual point of view, it can be interpreted as a behaviour that breaks conformal invariance of the boundary field theory \cite{Klebanov:1999tb, Witten:2001ua, Garay:2022szq}.  We can avoid this issue by considering a particular NED theory whose Lagrangian corresponds to\footnote{From the Power-law Lagrangian \eqref{PWLag} studied in \autoref{App:Power Law}, this corresponds to the conformal invariant case $p = 3/8$ in three dimensions, and $\gamma = -2^{-3/4}\lambda$~.}
\begin{align}
    \cL(\cS) = -2^{\frac14}\lambda |\cS|^{\frac34}~,
\end{align}
which is indeed conformal in three dimensions \cite{Hassaine:2007py}. The (traceless) energy-momentum tensor
\begin{align}
    T^\mu_\nu = \lambda |\cS|^{\frac34}\left(\delta^\mu_\nu - \frac{3F^{\mu\alpha}F_{\nu\alpha}}{\cS}\right)~,\end{align}
produces a Coulomb-like electric field
\begin{align}
    E(r) = \frac{\tq}{r^2}~,
\end{align}
where it has been chosen $\lambda =-|\tq|^{1/2}/2\lp$ and there is a black hole solution \cite{Cataldo:2000we} described by the metric function 
\begin{align}\label{CSsol}
    f(r) = \frac{r^2}{L^2} - m + \frac{2\tq^2}{r}~,
\end{align}
which indeed, is asymptotically AdS avoiding the aforementioned problem of the charged BTZ black hole. The solution has a proper spacetime singularity at $r = 0$ that can be seen from the quadratic invariant
\begin{align}
R_{\mu\nu\alpha\beta}R^{\mu\nu\alpha\beta} = \frac{12}{L^4} + \frac{12\tq^4}{r^6}~.   
\end{align}
Therefore, this solution shares more similarities with a Reissner--Nordstr\"{o}m-AdS black hole than the usual charged BTZ black hole.
The temperature of the solution 
\begin{align}
    T(x,\mu) = T_0\left(x - \tmu^2\right)~,\qquad \tmu^2 := \left(\frac{\mu}{2\pi}\right)^2\frac{1}{L}~.
\end{align}
From above, it is easy to see that the temperature is positive if $x \geq \tmu^2$ (for real chemical potential) and the specific heat at fixed chemical potential is
\begin{align}
    \cC = S^{(0)}-\V\tmu^2 = \frac{T}{T_0} \V~.
\end{align}
For vanishing charge, it reproduces the standard result $S^{(0)} > 0$ that guarantees local stability in the canonical ensemble.
The horizon radius can be computed analytically,
\begin{align}
    x_n = \frac{1}{n}+\tmu^2~.
\end{align}
In fact, the solution for the horizon radius is always real giving freedom to the value of $\mu$. This adds the possibility of  Wick rotate the chemical potential $\mu^2_{\rm E}= (i\mu)^2$ in the dual theory. Nonetheless, if we ask for a positive horizon radius, we have that $(\tmu_{E})^2 \leq n^{-1}$ for fixed $n$~. In the dual CFT, there is an upper bound for the imaginary chemical potential due to the non-analyticity of the free energy \cite{Belin:2013uta}. 

In order to highlight some features of this model, let us consider for a moment the difference of free energies, at same temperature, between thermal AdS$_{3}$, $I^E_{AdS}$, and the solution $I^E_{\rm C}$, 
\begin{align}
   \Delta I^E(\mu) \equiv I^E_{\rm C} - I^E_{AdS} = -\frac{\pi L x }{\lp}\frac{T_0}{T}\left(2x+\tmu^2\right)V_{\mathbb{H}^1}~,
\end{align}
showing no Hawking--Page phase transition for a real chemical potential. Instead, for imaginary chemical potential, we get
\begin{align}
    \Delta I^E(\mu_{\rm E}) \equiv I^E_{\rm C} - I^E_{AdS} = T_0 S^{(0)} \left(\frac{\tmu_{\rm E}^2-2x}{\tmu_{\rm E}^2+x}\right)~.
\end{align}
It clearly suggest that there could be a first order phase transition between thermal AdS and the black hole at critical temperature $T_c/T_0 = \frac{3}{2}\tmu_{\rm E}^2$~. This last point is radically different from the BTZ scenario where the Hawking--Page phase transition does not occur due to mass gaps between both configurations\footnote{In this case, it is possible to produce a Hawking--Page phase transition by considering contributions of conical deficits in the partition function \cite{Birmingham:2002ph, Kurita:2004yn, Eune:2013qs}.} \cite{Myung:2005ee}. Nonetheless, phase transitions in three-dimensional gravity coupled to NED seems to be dependant of the thermodynamical ensemble \cite{Hendi:2015wxa}, and the solution \eqref{CSsol} in the canonical ensemble does not shows phase transitions for real electric charge \cite{Cataldo:2020cxm}. 


As the solution is asymptotically AdS, we can compute the holographic quantities defined in the previous sections. Let us start with the magnetic response given in \eqref{magneticresponse},
\begin{align}
    k_n(\mu) = \frac{3R}{L}n\mu x_n~, 
\end{align}
and the conformal weight of the holographic twisted operators from \eqref{conformalweight} for imaginary chemical potential
\begin{align}\label{hn3}
    h_n(\mu) = \frac{\pi L}{\lp}\left[\left(n-\frac{1}{n}\right) + n\left(\frac{\mu_{\rm E}}{2\pi}\right)^4\right]~.
\end{align}
We can expand the conformal weights as done in \eqref{expansionh}, from which we read off the central charge $c$ by noting that 
\begin{align}
    \lim_{n\to 1}\lim_{\mu\to 0}\partial_n h_n(\mu) = \frac{2\pi L}{\lp}~.
\end{align}
The black hole entropy computed from the Wald formula is given simply by $S = \frac{\pi L}{\lp} V_{\mathbb{H}^1} x = \V x$~, and the Renyi entropy for an imaginary chemical potential is
\begin{align}
    S_n = \frac{\pi L}{\lp}\left(\frac{1}{2}\left(1+\frac{1}{n}\right)-\left(\frac{\mu_{\rm E}}{2\pi}\right)^2\right)V_{\mathbb{H}^1}~.
\end{align}
This result can be matched with the CFT computation of the Renyi entropy for free charged bosons in $d=2$ \cite{Belin:2013uta} 
by identifying $\left(\frac{\mu_{\rm E}}{2\pi}\right)^2 = \frac{|\mu_{\rm C}|}{2\pi}$ where $\mu_{\rm C}$ is the imaginary chemical potential of the conformal bosons. In this way, the rate in which the Renyi entropy decreases is dictated by the absolute value of the imaginary chemical potential in both theories. The generalised twist operators in a bosonic two-dimensional charged conformal field theory have conformal dimensions $h_{n}$ of the form $(\tilde{m}/n + \mu_{\rm E}/2\pi + \ell_{\tilde{m}})^2$ where $\tilde{m}$ and $\ell_{\tilde{m}}$ are constants. The linear term in $\mu_{E}$ can be eliminated by the $U(1)$ gauge freedom \cite{Belin:2013dva}, and the form of the conformal weights and Renyi entropy are exactly as \eqref{hn3} and \eqref{SnC} for small values of $\mu_{\rm C}$. 
The Renyi entropy limits are
\begin{align}
    &S_0 = \frac{\V}{2n}~, \quad  S_{\rm E} = \V(1 +\tmu^2)~,\quad  S_\infty = \frac{\V}{2}\left(1+2\tmu^2\right)~,\quad  S_2 = \V\left(\frac34 +\tmu^2\right)~,
\end{align}
and for large chemical potential
\begin{align}
\lim_{\mu\to\infty}S_n(\mu) = \V\tmu^2 + \frac{\V}{2n}(n+1)~,
\end{align}
that is, indeed, $n$-independant at the leading order. The Renyi entropy inequalities \eqref{ineq} can be checked analytically
\begin{align}
    \partial_n S_n ={}& -\frac{\V}{2n}~, && \partial_n\left(\frac{n-1}{n}S_n\right) = \frac{\V(1+n\tmu^2)}{n^3}~, \label{secineqcou}\\
    \partial_n\left((n-1)S_n\right) ={}& \frac{\V}{2}\left(1+\frac{1}{n^2} + 2\tmu^2\right)~, && ~\partial_n^2\left((n-1)S_n\right) = -\frac{\V}{n^3}~,
\end{align}
which are satisfied for all real values of $\mu$. Also notice, $\mu\partial_\mu S_n = 2\V\tmu^2$~, which is positive for real $\mu$~. Despite this, the inequalities restricts the value of $\mu^2$ at fixed $n$ when the chemical potential is imaginary while the positive horizon radius $\tmu_{\rm E} \leq n^{-1}$ ensures the inequalities to be satisfied. 

The thermal fluctuations are obtained using the prescription given in \autoref{Sec:ThermalFluctutations} in terms of the determinant of the matrix $D_{AB}$ and the Jacobian  $J$. In particular, the determinant,
\begin{align}
    {\rm det}D_{AB} = \frac{L((S^{(0)})^2-\lp L \V^2\mu^2)^2\left(\pi R(S^{(0)}-\lp L\V \mu^2)((S^{(0)})^2-\lp L \V^2\mu^2)-L\V^2\right)}{2\pi^2 R^2 (S^{(0)})^4\V^2}~,
\end{align}
that for the zero chemical potential goes as $3\log S^{(0)}$ coinciding with the value obtained by Carlip \cite{Carlip:2000nv} from considering thermal fluctuations on the torus partition function of a left- and right-moving CFT~. The Jacobian is
\begin{align}
    J = \frac{S^{(0)}\V}{(S^{(0)})^2-\lp L \V^2\mu^2}~,
\end{align}
and then, the corrected entropy becomes
\begin{align}
    S ={}& S^{(0)} - 3\log S^{(0)} +\log\left[\left((S^{(0)})^2-\lp L \V^2\mu^2\right)\left(((S^{(0)})-\lp L \V\mu^2)((S^{(0)})^2-\lp L\V^2 \mu^2)\right)\right]~,
\end{align}
plus terms that are independent of the entropy of the system. As we can see, the corrections contain a leading contribution of the form $-\nC\log S^{(0)}$ where we identify $\nC= 3$~. This kind of complicated expression for the fluctuation appear to be common for higher-dimensional black holes in open ensembles \cite{Ghosh:2021uxg}, and  are valid in the stability regime 
\begin{align}\label{CAcond}
    \cC \geq \frac{2\mu^2}{T}\frac{\V L}{2\pi R S^{(0)}}~.
\end{align}
This last inequality imposes a bound on the chemical potential given by 
\begin{align}\label{mubound}
    \tmu^4 -2\tmu^2\left(\frac{S^{(0)}}{\V} + \frac{L}{S^{(0)}}\right) + \left(\frac{S^{(0)}}{\V}\right)^2 \geq0~,
\end{align}
and \eqref{CAcond} recovers the standard $S^{(0)}\geq 0$ restriction for vanishing chemical potential.

Considering only the first correction, the  Renyi entropy renders
\begin{align}\label{SnC}
    S_n =  \frac{n}{n-1}\frac{\V}{2}\left(x_1^2-x_n^2\right) + \frac{n}{n-1}\nC\left(x_1 - x_n - x_1\log\V x_1 + x_n\log\V x_n\right)~,
\end{align}
with $\nC= 3$~. Again, the corrected charged Renyi entropy has the following limits
\begin{align}
    S_0 ={}& \frac{\V}{2n} - \nC\log\frac{\V}{n}~,\\ \label{SEC}
    S_{\rm E} ={}& \V(1+\tmu^2) - \nC\log\left(\V(1+\tmu^2)\right)~, \\ 
    S_\infty ={}& \frac{1}{2}(x_1 - x_\infty)(2\nC+\V(x_1 + x_\infty)) - \nC x_1\log \V x_1 + \nC x_\infty\log \V x_\infty~,\\
    S_2 ={}& \V\left(\frac34 -\tmu^2\right)+ \nC+\nC\left(1-2\tmu^2\right)\log\left(\V\left(\frac12-\tmu^2\right)\right) - 2\nC(1-\tmu^2)\log\left(\V(1-\tmu^2)\right)~,
\end{align}
where $x_1 = 1+\tmu^2$ and $x_\infty = \tmu^2$~. The large $\mu$ limit yields to
\begin{align}
    \lim_{\mu\to\infty}S_n(\mu) = \V\tmu^2 - \nC\log\left(\V\tmu^2\right)~,
\end{align}
which is again independent of $n$ at the leading order.

The thermal fluctuations generate a new subleading logarithmic divergence $\sim \log\V\tmu^2$ in the Renyi \eqref{SnC} as well the entanglement entropy \eqref{SEC}, and holds even in higher-dimensional Maxwell electrodynamics (see \autoref{App:Power Law}). As anticipated from \eqref{RenyiCorr}, this term is determined by the fact that the Bekenstein--Hawking entropy of the hyperbolic black hole acquires a logarithmic contribution due to fluctuations.  This entanglement entropy behavior has been observed in systems with non-smooth entangling regions \cite{Bueno:2015rda, Bueno:2015qya} or systems with SSB of a continuous symmetry due to low energy excitations of the ground state \cite{2011PhRvB..84p5134K, Metlitski:2011pr, Alba2013EntanglementSO, Kallin:2014oka, Frrot2016EntanglementEA, Casini:2019kex}. The term associated to the SSB diverges logarithmically with the system size, just as in \eqref{SEC}, and the coefficient in front is related to the number of Goldstones modes \cite{Metlitski:2011pr}. 
In \cite{Metlitski:2011pr}, the derivation of the subleading contributions was done by implementing a direct wave function calculation, and was proposed to have the form $\sim \log (L^{d-1}\alpha)$ where $L$ is the size of the system, and $\alpha$ gives information regarding the IR scales in the energy gaps of the theory which depends on intrinsic physical properties of the system, such as the total spin and spin-wave velocity. Then, we can associate the $\nC$ coefficient with number of Goldstone modes appearing in the dual theory, and $\alpha$ will depend on $\tmu$ and coefficients in $\V$ as can be seen from \eqref{SEC}. We further comment on this on \autoref{Sec:Conclusions}. 

\begin{figure}[t!]
\begin{center}
  \includegraphics[scale=0.45]{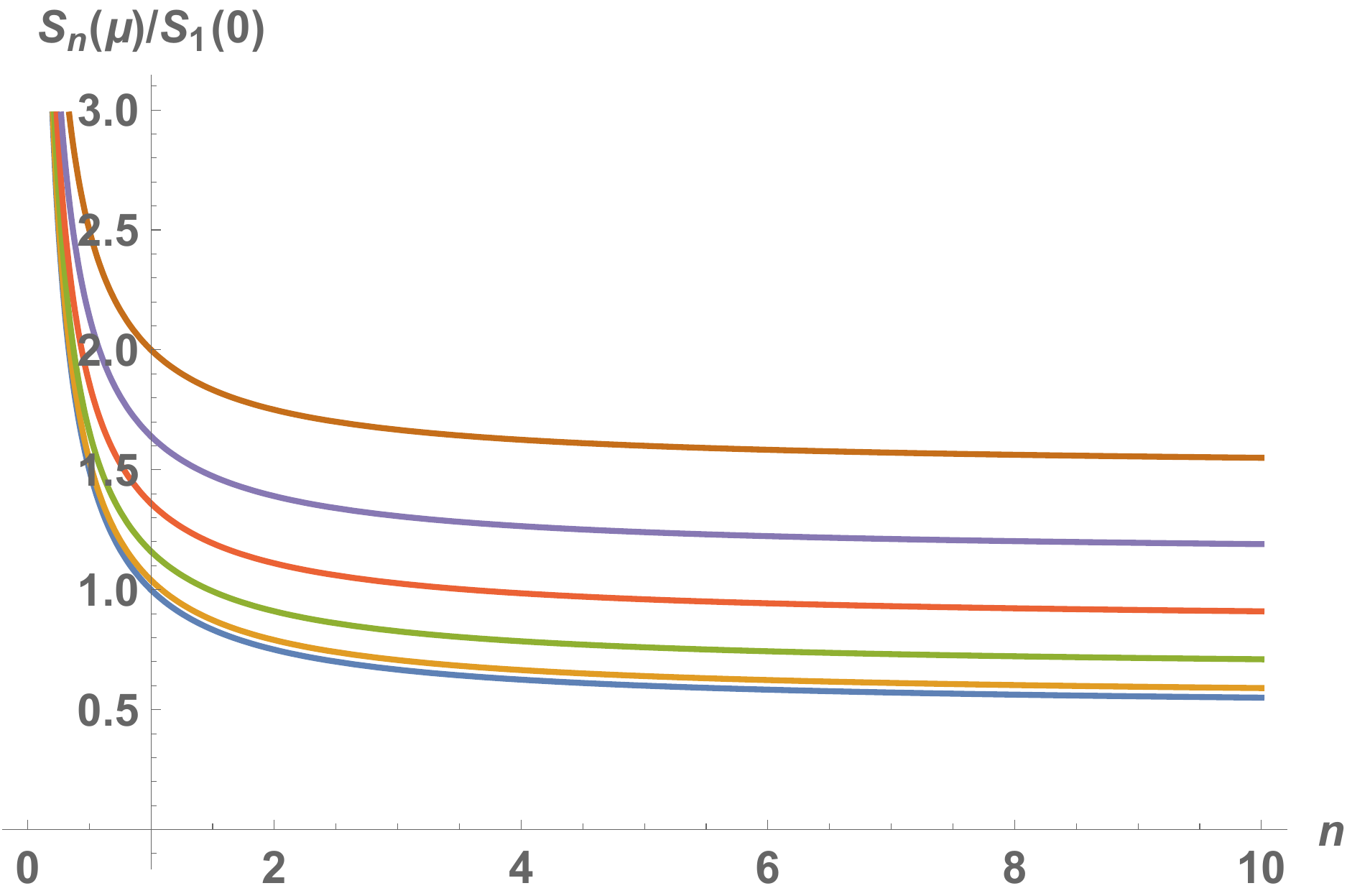}
  \includegraphics[scale=0.45]{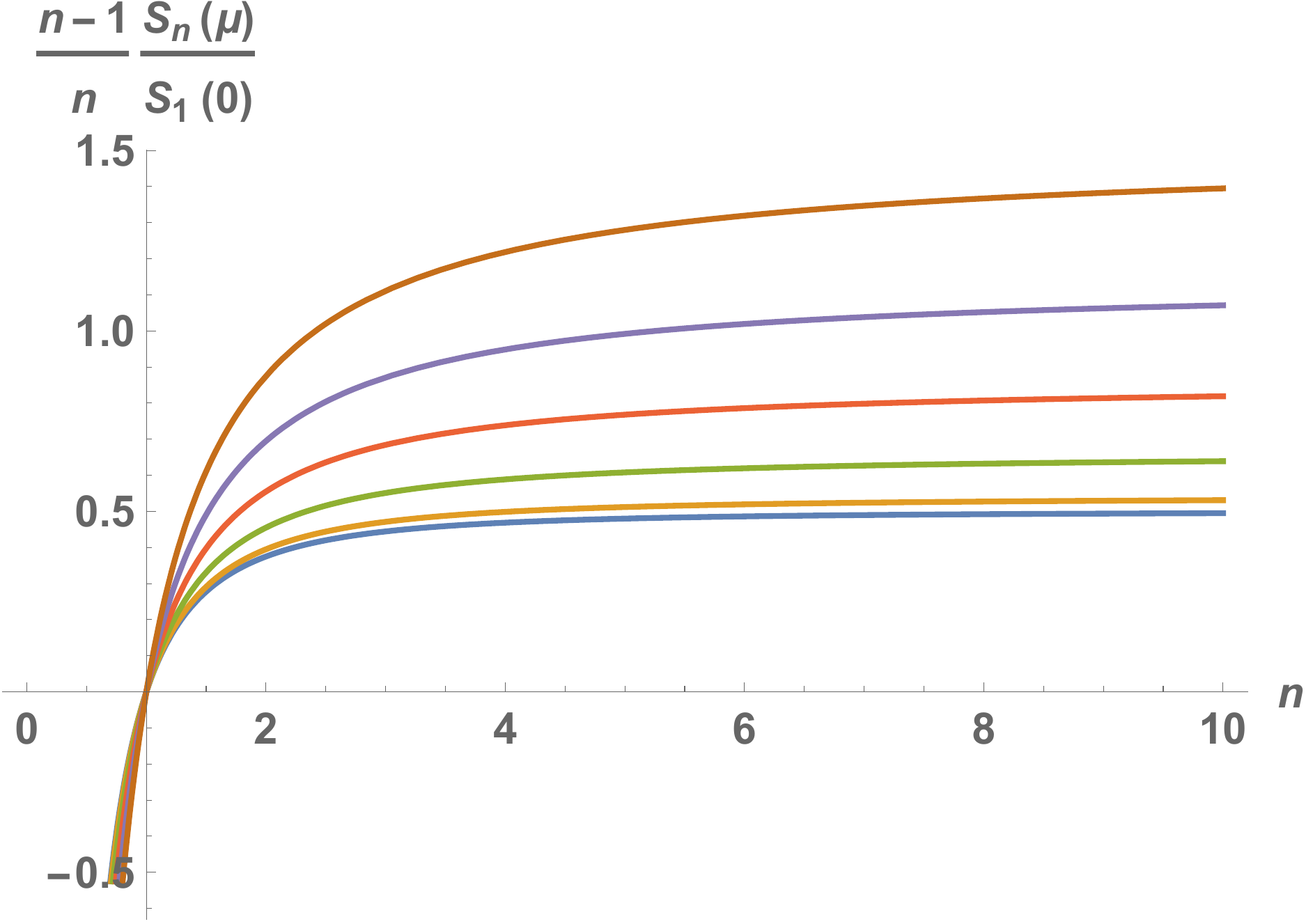}
  \captionof{figure}{Corrected Renyi entropy for three-dimensional gravity coupled to Coulomb sources with logarithmic corrections as a function of $n$ normalised by $S_1(0)$ on the left panel, and $\frac{n}{n-1}\frac{S_n(\mu)}{S_1(0)}$ on the right panel,  for different values of $\tmu$ with a UV cutoff so that $\V = 10^{40}$~. From bottom to top, the curves corresponds to $\tmu = 0,0.2,0.4,0.6,0.8,1$. }
  \label{fig:c1}
  \end{center}
\end{figure}

\begin{figure}[t!]
\begin{center}
  \includegraphics[scale=0.4]{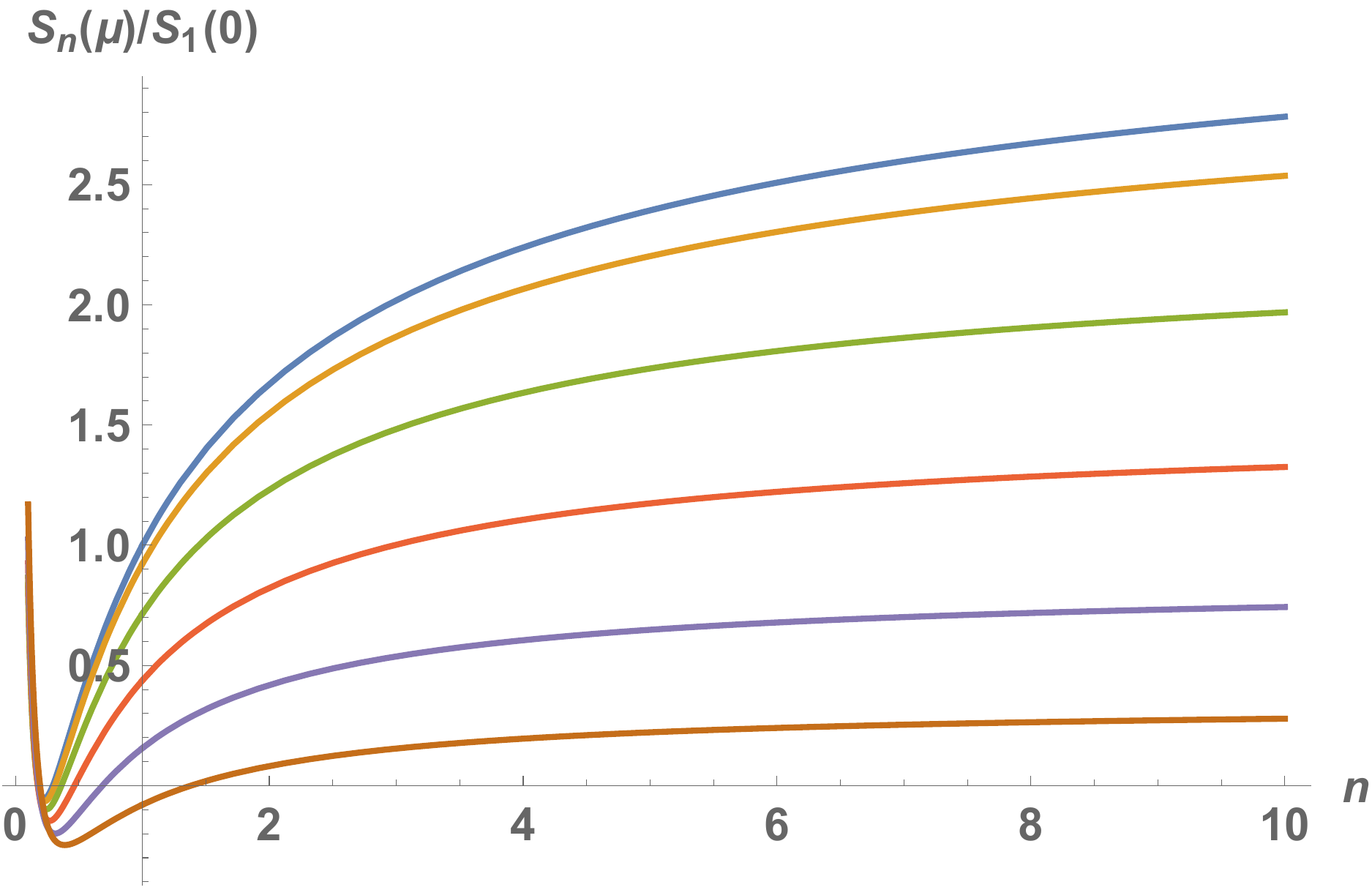}
  \includegraphics[scale=0.4]{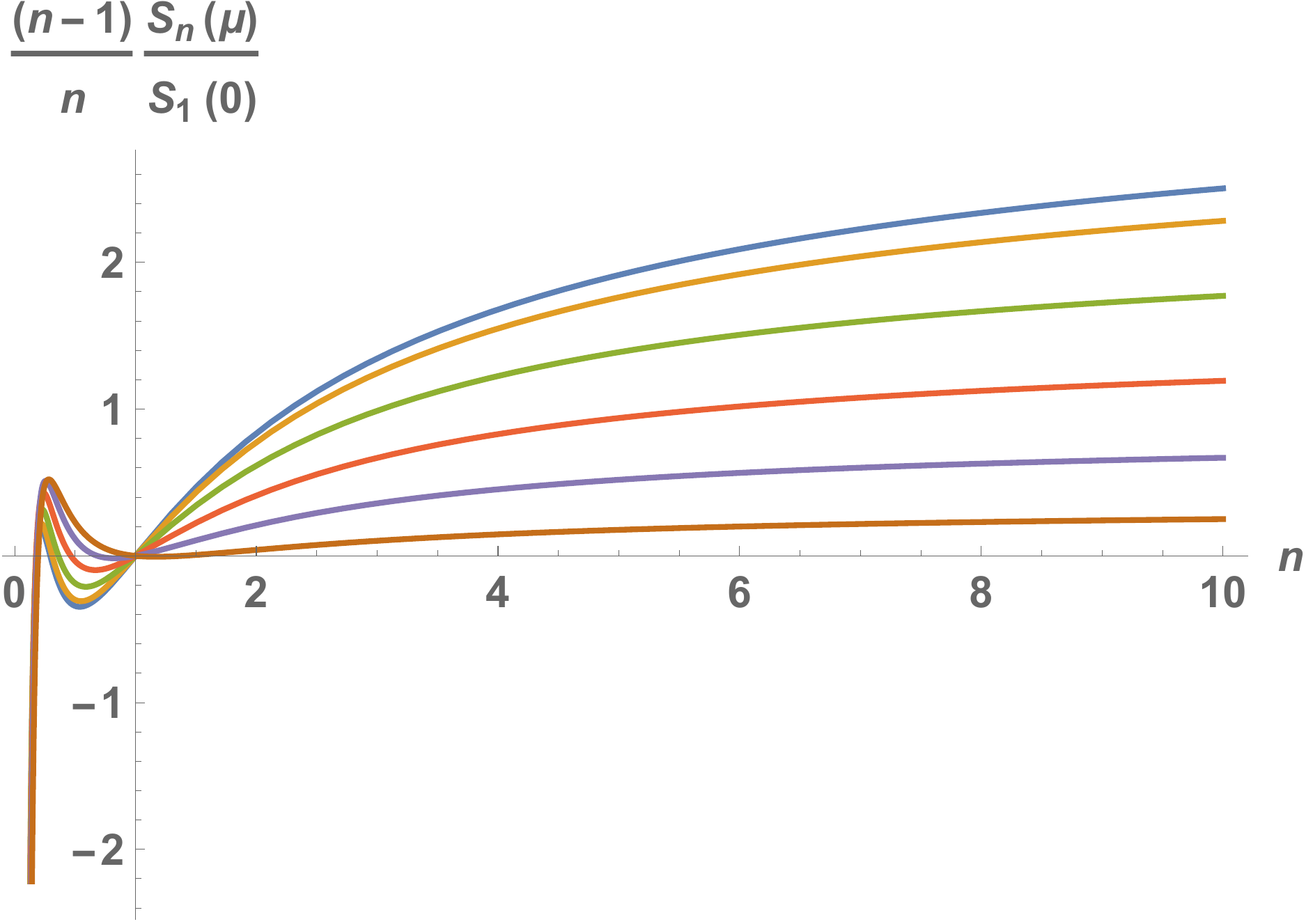}
  \captionof{figure}{Corrected Renyi entropy for three-dimensional gravity coupled to Coulomb sources with logarithmic corrections as a function of $n$ normalised by $S_1(0)$ on the left panel, and $\frac{n}{n-1}\frac{S_n(\mu)}{S_1(0)}$ on the right panel,  for different values of $\tmu$ with a UV cutoff so that $\V = 1$~. From top to bottom, the curves corresponds to $\tmu = 0,0.2,0.4,0.6,0.8,1$. }
  \label{fig:c4}
  \end{center}
\end{figure}

\begin{figure}[t!]
\begin{center}
  \includegraphics[scale=0.45]{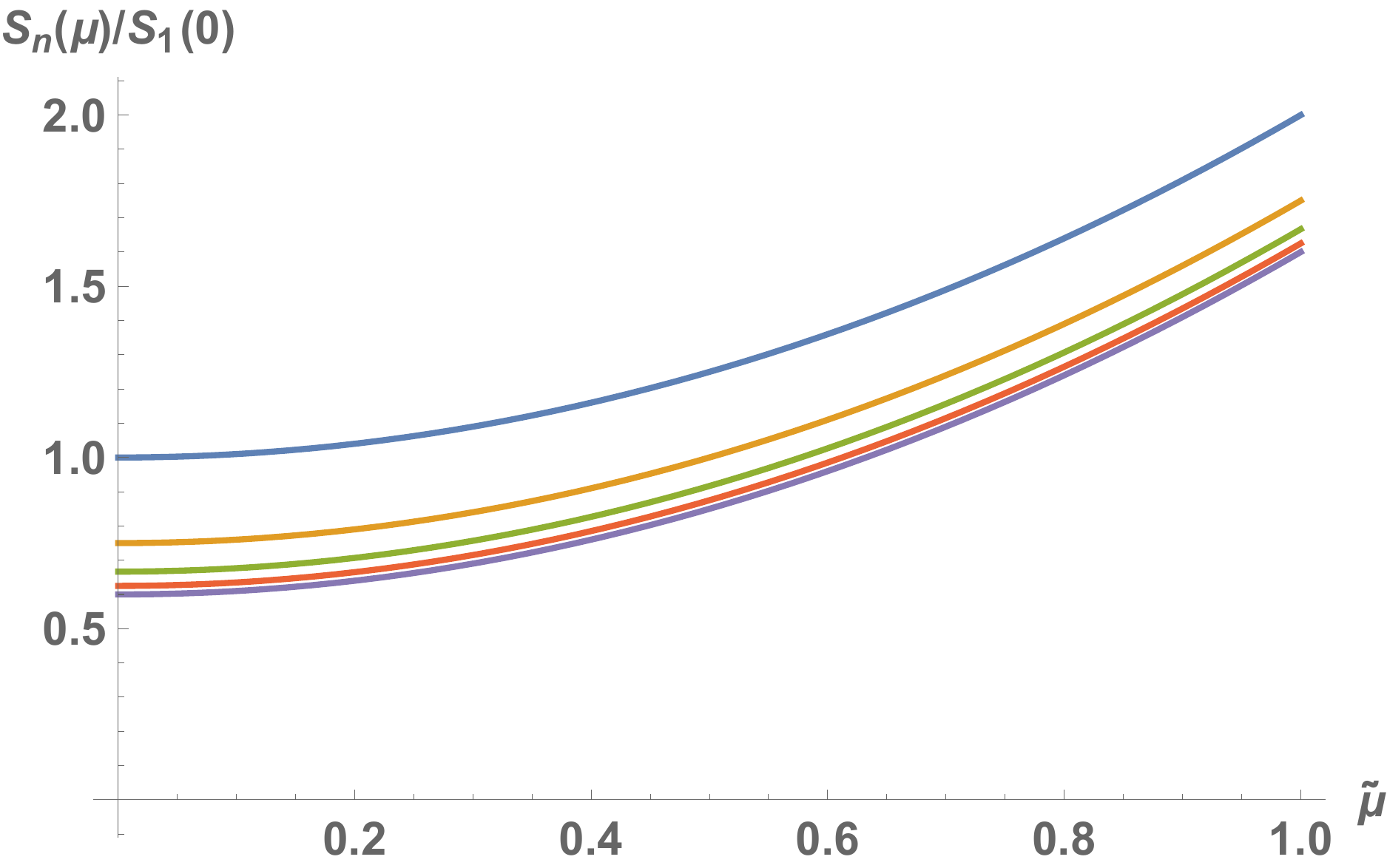}
  \includegraphics[scale=0.45]{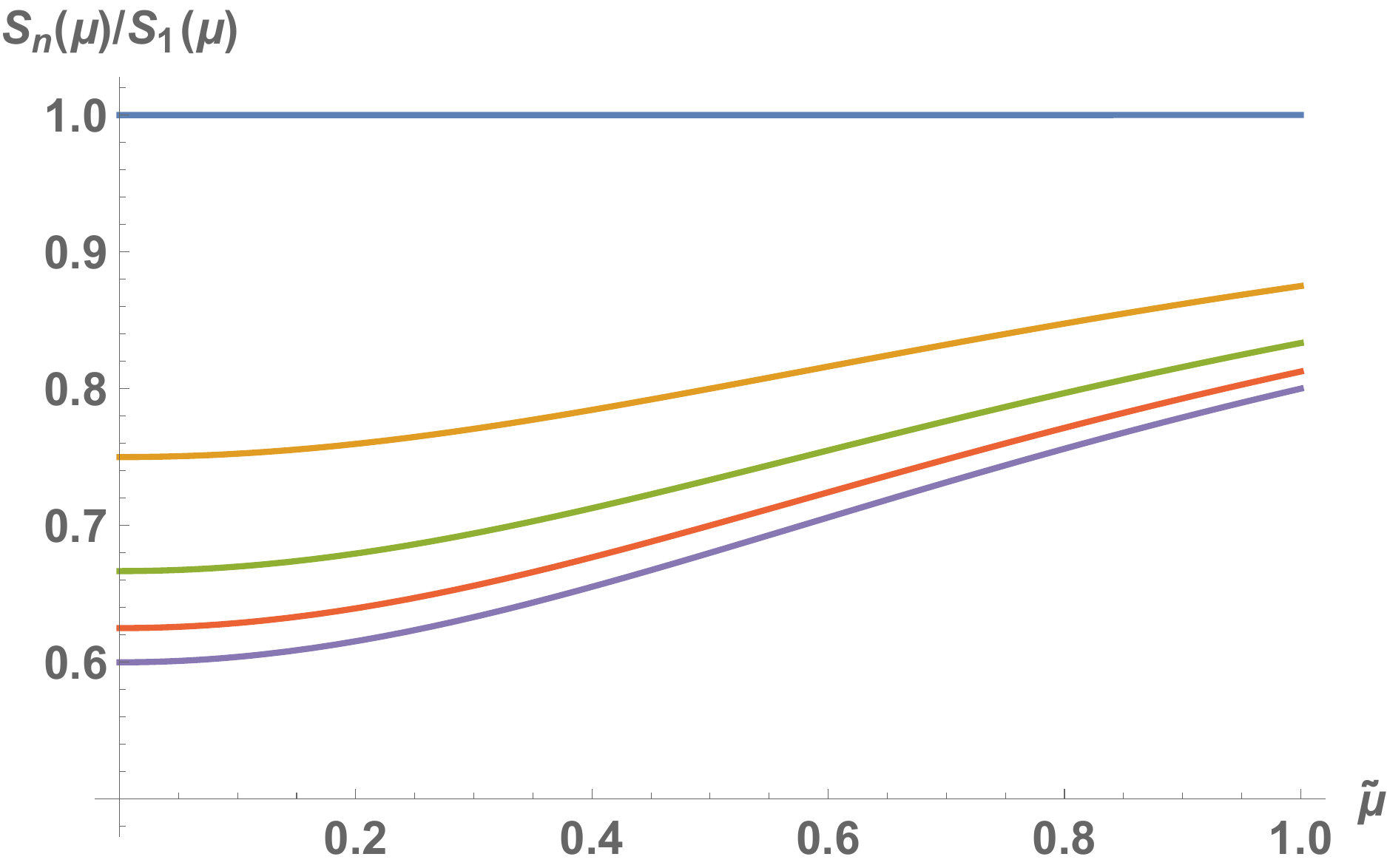}
  \captionof{figure}{Corrected Renyi entropy for three-dimensional gravity  coupled to Coulomb sources with logarithmic corrections as a function of $\tmu$ normalised by $S_1(0)$ on the left panel, and normalised by $S_1(\mu)$ on the right panel,  for different values of $\tmu$ with a UV cutoff such that $\V = 10^{40}$~. From top to bottom, the curves corresponds to $n = 1,2,3,4,5$ }
  \label{fig:c2}
  \end{center}
\end{figure}

\begin{figure}[t!]
\begin{center}
  \includegraphics[scale=0.45]{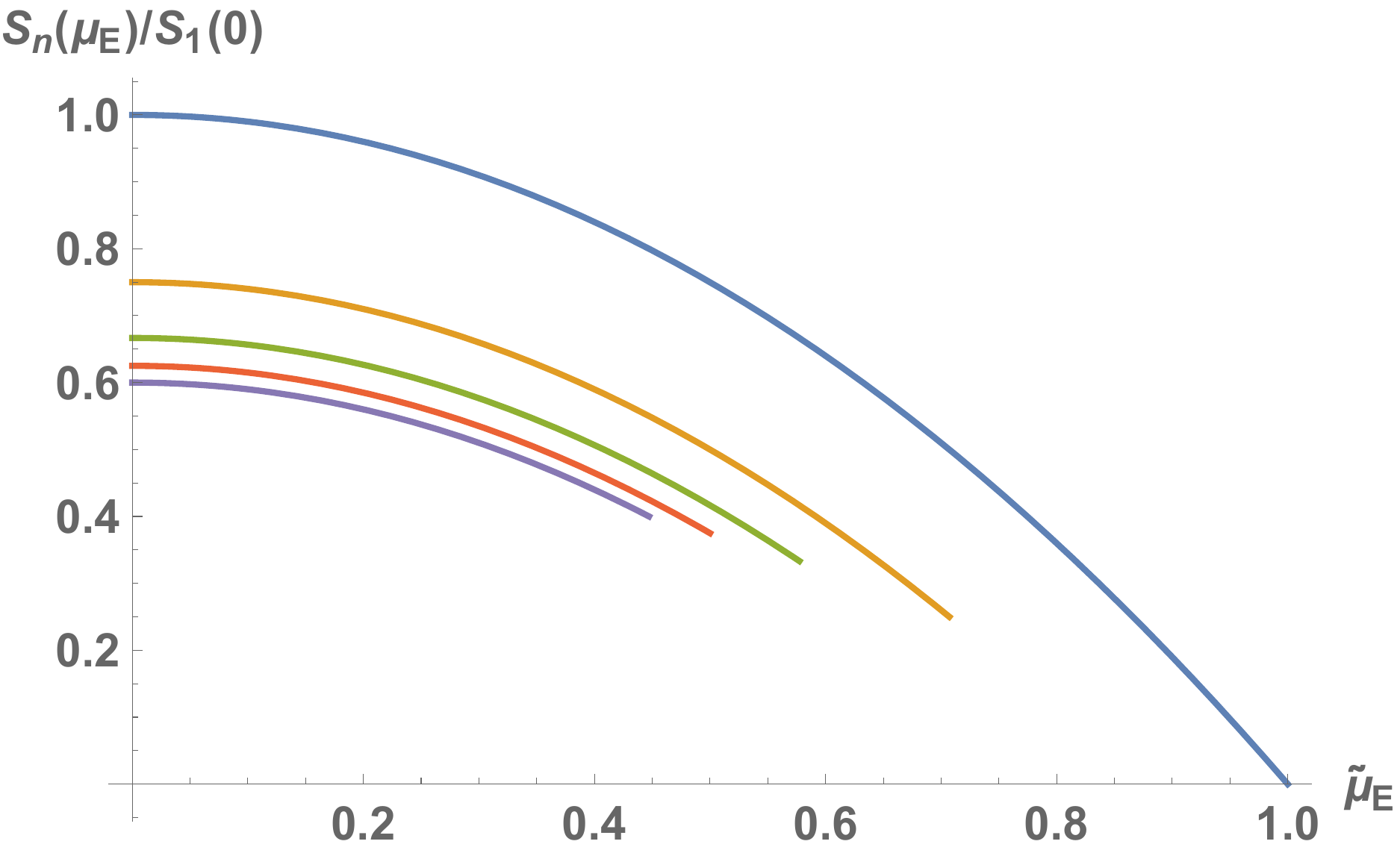}
  \includegraphics[scale=0.45]{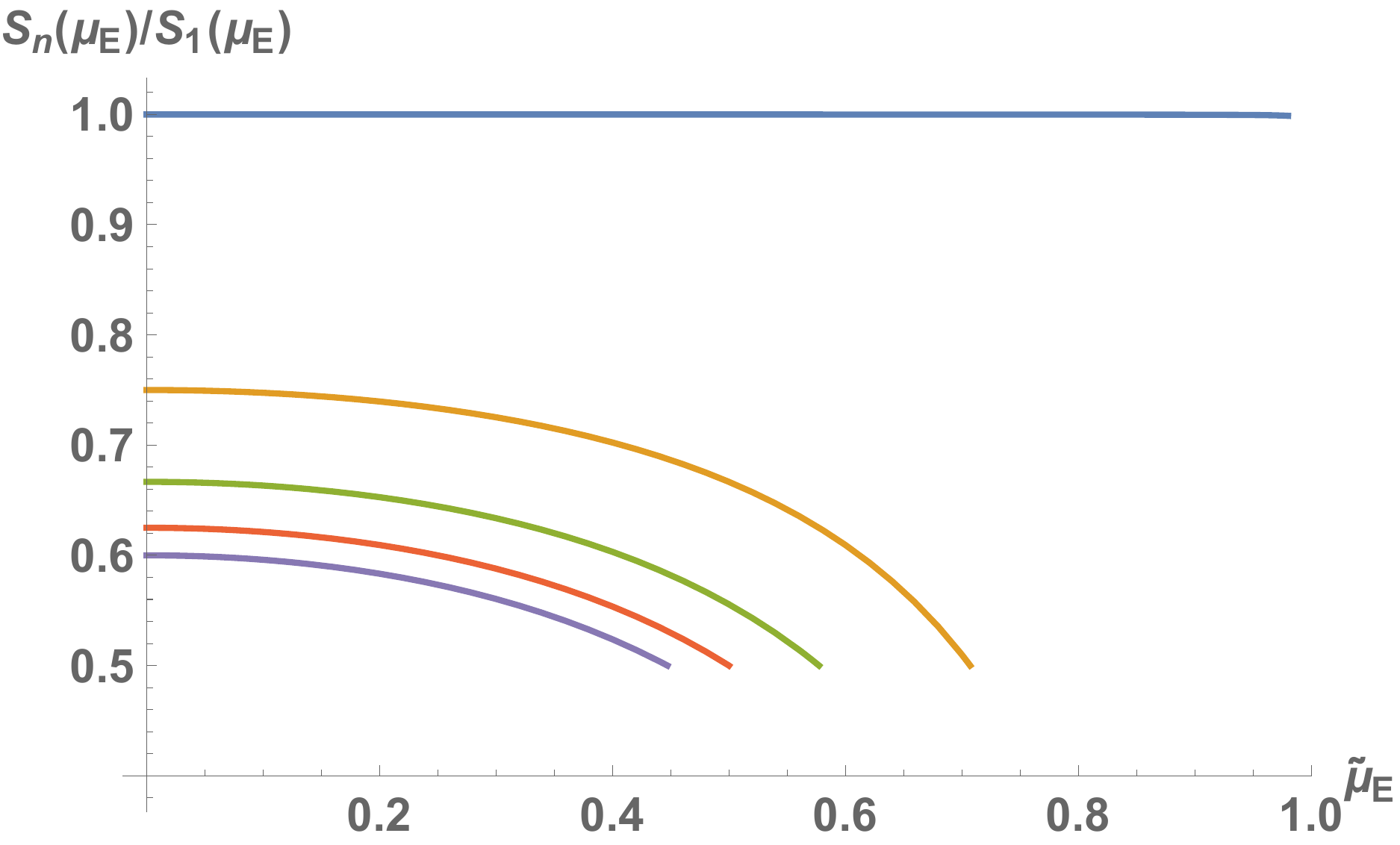}
  \captionof{figure}{Corrected Renyi entropy with imaginary chemical potential $\tmu_{\rm E} = i\tmu$ for Coulomb sources with logarithmic corrections normalised by the entanglement entropy with (non-)zero chemical potential in the (right) left panel. The UV cutoff is chosen such that $\V = 10^{40}$~. From bottom to top, the curves corresponds to $n=1,2,3,4,5$.}
  \label{fig:c3}
  \end{center}
\end{figure}

The derivatives of the modified Renyi entropy are
\begin{align}
    \partial_n S_n ={}& -\frac{(n-1)(2\nC n +\V(n-1)) + 2\nC(1+\tmu^2)\log\left(\frac{n}{n-1}(1+\tmu^2)\right)}{2n^2(n-1)^2}~, \\ \partial_n\left(\frac{n-1}{n}S_n\right) ={}&\frac{\V}{n^3}(1+n\tmu^2) - \frac{\nC}{n^3}\log\left(\frac{\V}{n}\left(1+n\tmu^2\right)\right) ~, \\
    \partial_n((n-1)S_n) ={}& \frac{\V}{2}\left(1+\frac{1}{n^2}+2\tmu^2\right) + \nC\left[\frac{n-1}{n} -\log\left(\V(1+\tmu^2)\right) + \tmu^2 \log\left(\frac{n^{-1}+\tmu^2}{1+\mu^2}\right)\right]~, \\ 
    \partial_n^2((n-1)S_n)) ={}& -\frac{1}{n^3}\left(\V -\frac{n\nC}{1+n\tmu^2} \right)~,
    \\ \mu\partial_\mu S_n ={}& 2\tmu^2\left(\V + \frac{\nC}{n-1}\log\frac{1+n\tmu^2}{n+n\tmu^2}\right)~, \label{muDc}
\end{align}
 which satisfy the Renyi inequalities \eqref{ineq} when the coefficient $\nC$ smaller than $\V$ as can be see from comparing \autoref{fig:c1} and \autoref{fig:c4}. The equation \eqref{muDc} shows that $\mu\partial_\mu S_n$ remains positive for $\nC\geq 0$ but now is $n$ dependant. As we mentioned earlier, the Renyi entropy for this solution shares many properties with the higher-dimensional Einstein-Maxwell case, even at the level of logarithmic corrections; for instance, in \autoref{fig:c1} the Renyi entropy approaches an asymptotic value that depends on the chemical potential. In \autoref{fig:c2}, we can see the growth of entangled pairs as $\tmu$ gets larger, and in \autoref{fig:c3}, the Renyi entropy decreases dramatically for imaginary chemical potential, and shows a maximum value in which the entropy is well-defined. Again, this resembles the free boson calculation where the analytic continuation is only valid within a finite window of values for the chemical potential. 
 Finally, the conformal weights receive the correction 
\begin{align}
    h_n(\mu) = \frac{\pi L}{\lp}\left[\left(n-\frac{1}{n}\right) + n\left(\frac{\mu_{\rm E}}{2\pi}\right)^4\right] - \frac{\nC}{V_{\mathbb{H}^1}}(1-x_n)n~,\quad \partial_n h_n(\mu)\big\rvert_{n\to1,\mu\to0} = \frac{2\pi L}{\lp} - \frac{\nC}{V_{\mathbb{H}^1}}~,
\end{align}
decreasing the value of the dual central charge.

\newpage

\clearpage

\section{Summary and Discussion}\label{Sec:Conclusions}
In this paper, we extended the holographic charged Renyi entropy to a bulk theory with general non-linear electrodynamics. Supplementing Einstein--Hilbert term with a Lagrangian density that depends on the Lorentz invariant $\cS=F_{\mu\nu}F^{\mu\nu}$~, allowed us to treat the thermodynamics of non-linearly charged black holes in a generic way without the need of specifying the model under consideration. With this, we constructed the holographic ingredients to compute the charged Renyi entropy: generalised charge \eqref{generalisedchargedensity}, magnetic response \eqref{magneticresponse}, and the twist operators  and their conformal weights \eqref{conformalweight}. 
We then considered thermal fluctuations in the grand canonical ensemble. The local stability of the system constraints the range of parameters in which the method to obtain the logarithmic corrections \cite{Mahapatra:2011si} is valid. The condition \eqref{Ccondition} is in fact, translated in terms of the energy-momentum tensor of the matter fields evaluated over the black hole horizon. The new corrected Renyi entropy \eqref{Sngeneral} can be easily separated between the ones coming from the Einstein--Hilbert term and the ones given by the gauge sector. This comes in handy to recover the uncharged case \cite{Mahapatra:2016iok} given by turning off the chemical potential. The Renyi inequalities \eqref{ineq} are satisfied as long as the thermal entropy of the system is positive, constraining the value of the coefficient $\nC$ in terms of the entropy.

We interpret these new contributions by studying specific examples; the Coulomb source in 2+1 dimensions \autoref{Sec:Coulomb} and Maxwell theory in $d\geq3$ in \autoref{App:Power Law}. In both cases, the structure of the subleading contributions to the Renyi entropy have the flavour of coming from spontaneous symmetry breaking of some continuous symmetry in the CFT. If we follow the holographic dictionary, the duality matches UV excitation in one side to IR effects in the other. This means that the corrections that we have found, from the gravity side, corresponds to high energy fluctuations beyond the classical saddle which from the boundary field theory point of view, are low energy excitations. In fact, the SSB contribution to the entanglement entropy does not modifies its UV structure \cite{Metlitski:2011pr}, and therefore it cannot be seen holographically from the classical saddle in the AdS partition function \cite{Jeong:2022zea,Park:2022oek}. A proposal for quantum corrections of holographic entanglement entropy was given in \cite{Faulkner:2013ana}, were it is claimed to be dual to the relative entropy between bulk regions disconnected by the Ryu--Takayanagi surface. In the Klebanov--Strassler theory at large $N$, the corrections, which are zero order in the gravity coupling, are associated to a spontaneously broken global symmetry due to bulk massless excitations, and have the same subleading logarithmic structure as the one presented in \autoref{Sec:ThermalFluctutations}. Similar corrections can be also found in other holographic theories, as in \cite{Fujita:2009kw}, by considering 1-loop contributions to the gravity partition function. 

Considering thermal fluctuations leads to new terms with the same structure which allow us to identify the $\nC$ coefficient with the number of Goldstone modes in the dual theory. This matching, gives the possibility to study quantum correlations in theories with SSB and Maxwell fields by means of holography beyond the saddle point as pointed out in \cite{Casini:2019kex}. It is important to remark that the Renyi inequalities ensure unitarity and stability of the boundary theory constraining the holographic dictionary \cite{Nakaguchi:2016zqi}. This restricts the possible geometries to be considered in the partition function, see \cite{Maloney:2007ud} for a discussion in three-dimensions.
Furthermore, the consideration of extra Maxwell fields in a system with SSB leads to a double logarithmic divergence $\sim \log\log L$ in the quantum mutual information, which only depends on the system length \cite{Casini:2019kex, Huerta:2022cqw}. The mutual information is a quantity measuring correlation between subsystems, and is constructed in terms of the entanglement entropy in such a way that is independent of renormalisation schemes. In our case, a double logarithmic structure in the entanglement entropy appears for odd dimensional bulk due to the subleading contributions in the divergent volume $\V$ \cite{Myers:2010xs} which are of the form $\sim \log L$~, which also holds in Maxwell theory in any dimensions as can be seen in \eqref{SEMax}~. This indicates the possibility of obtaining the double logarithmic structure in the quantum mutual information by considering fluctuations generating a logarithmic correction in the black hole entropy. We plan to refine and present these ideas in a separate work. 

The breaking of continuous symmetries should also influence the gravity theory. As a matter of fact, this should be seen as some ``quantum'' backreaction in the geometry due to Gaussian fluctuations in the thermal partition function \cite{Kazakov:1993ha}. The backreaction can be modeled by adding the Barvinsky--Vilkovisky--de Witt non-local effective action \cite{Vilkovisky:1984st, Barvinsky:1983vpp, DeWitt:1985sg} that, indeed, produces the logarithmic corrections on the black hole entropy \cite{Calmet:2021lny, Xiao:2021zly, Pourhassan:2022auo, Pourhassan:2022opb}. The deformations can break the original black hole symmetries generating terms of the form of \eqref{btzlog}, as can be seen in \cite{Pourhassan:2022auo}.  This could lead to model breaking of conformal invariance in the AdS boundary when the quantum corrections dominate over the saddle, as could be for final stages of black hole due to evaporation.
Indeed, the backreaction of the energy fluctuations in the geometry has been shown to break local symmetries of the solution, such as the spherical symmetry of the Schwarzschild black hole \cite{Kazakov:1993ha}. This breaking of local symmetries in the bulk should correspond to breaking the CFT global spherical symmetry being a concrete realisation of our proposal.  We plan to refine this idea in a future work.

We also noticed that the fluctuations modify the conformal weights of the twist operators, which leads to a subleading modification of the numerical value of the central charge. This is in contrast with the magnetic response that does not suffer a direct adjustment due to the presence of the $\nC$, leaving the structure coefficients associated with the $\langle JJ\rangle$ 
 and $\langle TJJ\rangle$ correlators unmodified, which encaptures universal features of the CFT \cite{Belin:2013dva, Bueno:2022jbl}, but modifying the ones of the stress-tensor two point function. In order to see modifications of the magnetic response one would have to consider the aforementioned modifications of the black hole geometry in order to modify the thermodynamic quantities such as the electric charge.

Another interesting result was obtained in \autoref{Sec:Coulomb} for a particular case of NED in three-dimensional gravity. In the standard description of 2+1 gravity coupled to Maxwell, the BTZ black holes acquires a logarithmic contribution to the metric due to the presence of the electric charge. This logarithmic term modifies the asymptotic behavior of the solution being not asymptotically AdS any longer. Then, the holographic charged Renyi entropy does not reproduce the expected free theory in the bulk as in dimensions.
Instead, to avoid this problem, we considered the Coulomb source term that produces a charged black hole with global AdS in the asymptotic region \cite{Cataldo:2000we}. In this way, the dual field theory is conformally invariant which allows to properly interpret the bulk quantities using the holographic dictionary. We showed that the structure for the conformal weights of the twist operators, and the charged Renyi entropy, is the same as the one found in \cite{Belin:2013dva} for charged free bosons in 1+1 dimensions. This offers the possibility of a complete correspondence between these two models. However, it also brings a few points that are need to be investigated in order to have a concrete statement about the duality:

\paragraph{$\diamond$} From a classical perspective, the solution presented should be geodesically complete. This has not been clarified in the previous literature, and it corresponds to an important issue when understanding the isometries of the boundary theory.

\paragraph{$\diamond$} The asymptotic behavior of the bulk fields would give insight about the sources and vacuum expectation values of the boundary operators. For this, we require to apply holographic renormalisation techniques \cite{deHaro:2000vlm} that also would allow us to write down the Euclidean action and holographic stress tensor, in order to fully characterise the boundary field theory. 

\paragraph{$\diamond$} It would be interesting to see if the three-dimensional NED theory arises as a sector of String theory. There are two possible ways of how to obtain this Lagrangian: As coming from a conformal limit of an effective description of a theory that interacts with matter, or as a Kaluza--Klein reduction of higher-curvature gravity that would correspond to IR limits of String/M theory \cite{Gibbons:2000xe}.
\\

We hope that these questions and issues can be examine with more detail in the future. 
\section*{Acknowledgments}
We would like to thank A. Das, A. Donos, S. Hosseini, V.G. Matyas, R. Olea, and P. Sundell for enlightening discussions and comments on various topics related to this manuscript. 
The work of {\sc Gah} is funded by {\sc Becas Chile} (ANID) Scholarship N$^{\rm o}~ 72200271$. 
The work of {\sc FD}~is supported by {\sc Beca Doctorado nacional} 
(ANID) 2021 Scholarship N$^{\rm o}~21211335$ and ANID/ACT210100 Anillo Grant ``{\sc Holography and its applications to High Energy Physics, Quantum Gravity and Condense Matter Systems}''. 

\begin{appendix}
\section{Power-law Maxwell electrodynamics}\label{App:Power Law}
Consider the non-linear Lagrangian\footnote{See \cite{Jing:2011vz} and reference therein for holographic applications of the Power-law Maxwell Lagrangian.},
\begin{align}\label{PWLag}
    \cL = -2\gamma F^{2p}~,
\end{align}
where $\gamma$ is a real constant that, in principle, can depend on the electric charge of the theory, but we will consider in this Appendix both quantities to be independent. 
The energy-momentum tensor
\begin{align}
    T^\mu_\nu = -2\gamma \left(\delta^\mu_\nu - 4p \frac{F^{\mu\alpha} F_{\nu\alpha}}{F^2}\right)F^{2p}~,
\end{align}
generates an electric field and potential
\begin{align}
    E(r) = \frac{Q}{r^\delta}~, \qquad \phi(r) = \frac{Q}{(1-\delta)r^{\delta-1}}~,
\end{align}
where
\begin{align}
    \delta = \frac{d-1}{2p-1}~,\qquad Q = \left(\frac{(-1)^{p+1}\tq}{2^p p\gamma}\right)^{\frac{1}{2p-1}}~,\qquad \frac{\mu_{\rm p}}{2\pi L} = \frac{Q}{(1-\delta)r_h^{\delta-1}}~.
\end{align}
Notice that when $p = \frac{d+1}{4}$~, the electric field always has the Maxwell form  and the Lagrangian becomes conformal leaving the energy-momentum traceless \cite{Hassaine:2007py}. The $rr$-component of the energy-momentum tensor is
\begin{align}
    T_r^r = (-1)^{p}(2p-1)\gamma 2^{p+1}\frac{Q^{2p}}{r^{2\delta p}}~,
\end{align}
and that evaluated at the horizon gives
\begin{align}
    T^r_r(x,\mu_{\rm p}) = \gamma(-1)^p \left(\frac{\tmu_{\rm p}}{x L}\right)^{2p}~,\qquad \tmu^{2p}\equiv (2p-1)2^{p+1}\left(\frac{(1-\delta)\mu_{\rm p}}{2\pi L}\right)^{2p}~.
\end{align}
Its derivative 
\begin{align}
    \partial_x T_r^r = \gamma(-1)^{p+1}\left(\frac{2p}{x} \right)\left(\frac{\tmu_{\rm p}}{x L}\right)^{2p}~,
\end{align}
is positive for $p$ odd and satisfies the stability condition in the canonical ensemble~. In the case of $p$ even,  must be analysed to check the temperature regime of stability, see \cite{Gonzalez:2009nn} for a discussion.  
In the case of even $p = 2m$~, we get an upper bound for the chemical potential
\begin{align}
    d + \frac{d-2}{x^2}\geq \frac{2\lp^{d-1}L^2}{\delta}\gamma \left(\frac{\tmu_{\rm p}}{L x}\right)^{4m}~.
\end{align}

The charge dependent sector of the metric function becomes
\begin{align}
    \cQ(r) = -\frac{2\gamma(d-1)(-2)^{p}Q^{2p}}{\delta(\delta-1)}\frac{1}{r^{\delta-1}}~,
\end{align}
such that, in terms of $\tmu_{\rm p}$, the specific heat at fixed chemical potential
\begin{align}
\cC = (d-1)S^{(0)}\left[\frac{dx^2 +k(d-2) + \frac{2(-1)^p\gamma\lp^{d-1}L^2 x^2 }{d-1}\left(\frac{\tmu_{\rm p}}{x L}\right)^{2p}}{dx^2 - k(d-2) + \frac{2(2p-1)(-1)^p\gamma\lp^{d-1}L^2 x^2}{d-1}\left(\frac{\tmu_{\rm p}}{x L}\right)^{2p}}\right]~,
\end{align}
and temperature
\begin{align}
    T =  \frac{T_0}{2x_n}\left[dx_n^2 -(d-2) +(-1)^{p}\gamma\frac{2\lp^{d-1}}{d-1} \left(\frac{\tmu_{\rm p}}{L} \right)^{2p}\frac{1}{x_n^{2p-2}}\right]~,
\end{align}
where $x_n$ is the largest solution to
\begin{align}
 dn x_n^{2p} -n(d-2)x_n^{2p-2} -2x_n^{2p-1} 
 + n(-1)^{p}\gamma\frac{2\lp^{d-1}}{d-1} \left(\frac{\tmu_{\rm p}}{L} \right)^{2p} =0~.
\end{align}
Analytic solutions for $x_n$ can be found for several values of $p$~. Particularly, for the Maxwell case, $p=1$, the equations is quadratic, and when the theory becomes conformal the equation becomes
\begin{align}
  dnx^{\frac{d+1}{2}} - 2x_n^{\frac{d-1}{2}}-n(d-2)x_n^{\frac{d-3}{2}} + (d-2)\tmu^{\frac{d+1}{2}} = 0~,
\end{align}
which is quadratic in $d=3$ and corresponds to Maxwell theory~.

Considering the pure Maxwell, which corresponds to $p=1$ with $\gamma = \frac{\ell_\ast^2}{16\lp^{d-1}}$~, and following the conventions of \cite{Belin:2013uta}, the Lagrangian becomes
\begin{align}
\cL=- \frac{\ell_\ast^2}{8 \lp^{d-1}}  F_{\mu\nu}F^{\mu\nu}~,
\end{align}
and the theory admits a charged solution with a hyperbolic horizon when $d>2$ with metric function 
\begin{align}
    f(r) = \frac{r^2}{L^2}-1-\frac{m}{r^{d-2}}+\frac{q^2}{r^{2(d-2)}}~,
    \end{align}
and corresponding gauge field 
\begin{align}\label{Amax}
        A = \left(\sqrt{\frac{2(d-1)}{(d-2)}}\frac{Lq}{R\ell_\ast r^{d-2}} - \frac{\mu_{\rm M}}{2\pi R}\right)dt~.
\end{align}
where $\mu_{\rm M}$ is chosen such that the gauge field vanishes at the horizon,  namely
\begin{align}
    \mu_{\rm M} = 2\pi\sqrt{\frac{2(d-1)}{(d-2)}}\frac{Lq}{\ell_\ast r^{d-2}_{h}}~.
\end{align}
There is a black hole horizon at $r_{h}$ and the mass parameter can be written as a function of the horizon,
\begin{align}
    m = \frac{r_h^{d-2}}{L^2}(r^2_h - L^2) + \frac{q^2}{r_h^{d-2}}~,
\end{align}
As we mentioned in sec
 the full solution is parametrised by the AdS radius $L$, the electric charge $q$ and the horizon $r_{h}$,

\begin{align}
    f(r) = \frac{r^2}{L^2} - 1 +\frac{q^2}{r^{2d-4}}-\left(\frac{r_h}{r}\right)^{d-2}\left(\frac{r_h^2}{L^2}-1+\frac{q^2}{r_h^{2d-4}}\right)~.
\end{align}
The black hole has an associated temperature given by
\begin{align}
    T = \frac{T_0}{2}L f'(r_h) = \frac{T_0}{2}\left[d\frac{r_h}{L}-(d-2)\frac{L}{r_h}\left(1+\tmu_{\rm M}^2\right)\right]~, \qquad \tmu_{\rm{M}} \equiv \sqrt{\frac{d-2}{2(d-1)}}\left(\frac{\mu_{\rm M}\ell_\ast}{2\pi L}\right),
\end{align}
and entropy given by the Wald formula evaluated over the horizon, 
\begin{align}
    S^{(0)} = \V \left(\frac{r_h}{L}\right)^{d-1}~.
\end{align}
The horizon radius in terms of the Renyi parameter is
\begin{align}
x_n = \frac{1}{dn}\left(1 + \sqrt{1+d(d-2)(1+\tmu_{\rm M}^2) n^2}\right)~.
\end{align}
The heat capacity has been computed in \cite{Cai:2004pz}, in agreement with \eqref{specificheat}, and shown to be always stable. 
Considering the first order corrections in \eqref{LogS} can be used and gives
\begin{align}\label{Scorr}
    S = S^{(0)} - \nC_{\rm M} \log S^{(0)} + {\rm const.}~,\qquad \nC_{\rm M}  = 1~,
\end{align}
and the coefficient $\nC_{\rm M} $ has been found in \cite{Gour:2003jj} for arbitrary dimensional AdS RN in the (grand) canonical ensemble.
Using the result into the Renyi entropy
\begin{align}\label{RenyiT}
    S_n = S_n^{(0)} + S_n^{(1)}~,\
\end{align}
where the first order $S^{(0)}_n$ was found in \cite{Belin:2013uta},
\begin{align}
    S^{(0)}_n = \frac{\V}{2} \frac{n}{n-1}\left[\left(1 + \tmu^2\right)(x_1^{d-2}-x_n^{d-2}) + x_1^d - x_n^d\right]~,
\end{align}
and the first correction
\begin{align}\label{C1}
    S^{(1)}_n ={}& - \frac{\nC_{\rm M} }{2}\left(\frac{n}{n-1}\right)\int^{x_1}_{x_n}  \left(d + (d-2)(1+\tmu_{\rm M}^2) x^{-2}\right)\log\left(\V x^{d-1}\right)dx \nonumber \\ ={}& \frac{n}{n-1}\frac{\nC_{\rm M}}{2}(d-2)(1+\tmu_{\rm M}^2)\left(\frac{\log\left(\V x_1^{d-1}\right)+(d-1)}{x_1} - \frac{\log\left(\V x_n^{d-1}\right)+(d-1)}{x_n}\right)~.
\end{align}
The resulting Renyi entropy \eqref{RenyiT} has limits
\begin{align}
    S_{0} {}&= \frac{\V}{2}\left(\frac{2}{d}\right)^d\frac{1}{n^{d-1}} - \nC_{\rm M} \left[\log\left(\left(\frac{2}{d n}\right)^{d-1}\V\right)-(d-1)\right]~, \\[5mm] \label{SEMax} S_{\rm E} {}&= \frac{\V}{2}\frac{(d-2)x^{d-2}_1}{dx_1-1}\left(\frac{dx
    _1^2}{d-2}+1 + \tmu_{\rm M}^2\right) -\nC_{\rm M} \log\left(\V x_1^{d-1}\right)~, \\[5mm] S_\infty {}&= \frac{\V}{2}\left[\left(1+\tmu_{\rm M}^2\right)\left(x_1^{d-2}-x_\infty^{d-2}\right) + x_1^d - x_\infty^d\right] \nonumber \\ {}& +\frac{\nC_{\rm M}}{2}(d-2)(1+\tmu_{\rm M}^2)\left(\frac{\log\left(\V x_1^{d-1}\right)+(d-1)}{x_1} - \frac{\log\left(\V x_\infty^{d-1}\right)+(d-1)}{x_\infty}\right)~, \\[5mm] S_2 {}&= \V\left[\left(1+\tmu_{\rm M}^2\right)\left(x_1^{d-2}-x_2^{d-2}\right) + x_1^d - x_2^d\right]  \nonumber \\ {}& +\nC_{\rm M}(d-2)(1+\tmu_{\rm M}^2)\left(\frac{\log\left(\V x_1^{d-1}\right)+(d-1)}{x_1} - \frac{\log\left(\V x_2^{d-1}\right)+(d-1)}{x_2}\right)~.
\end{align}

\begin{figure}[t!]
\begin{center}
  \includegraphics[scale=0.4]{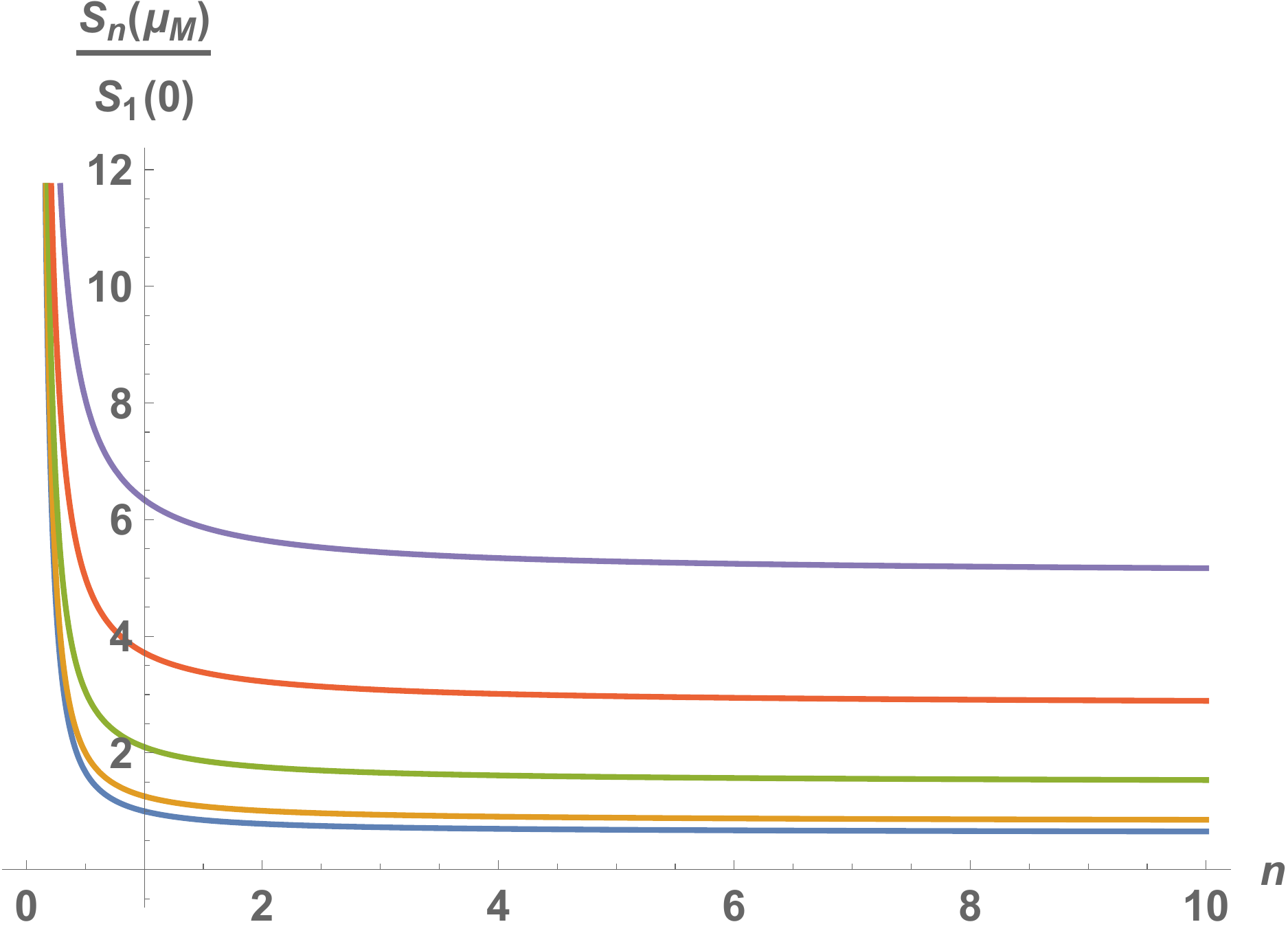}
  \includegraphics[scale=0.4]{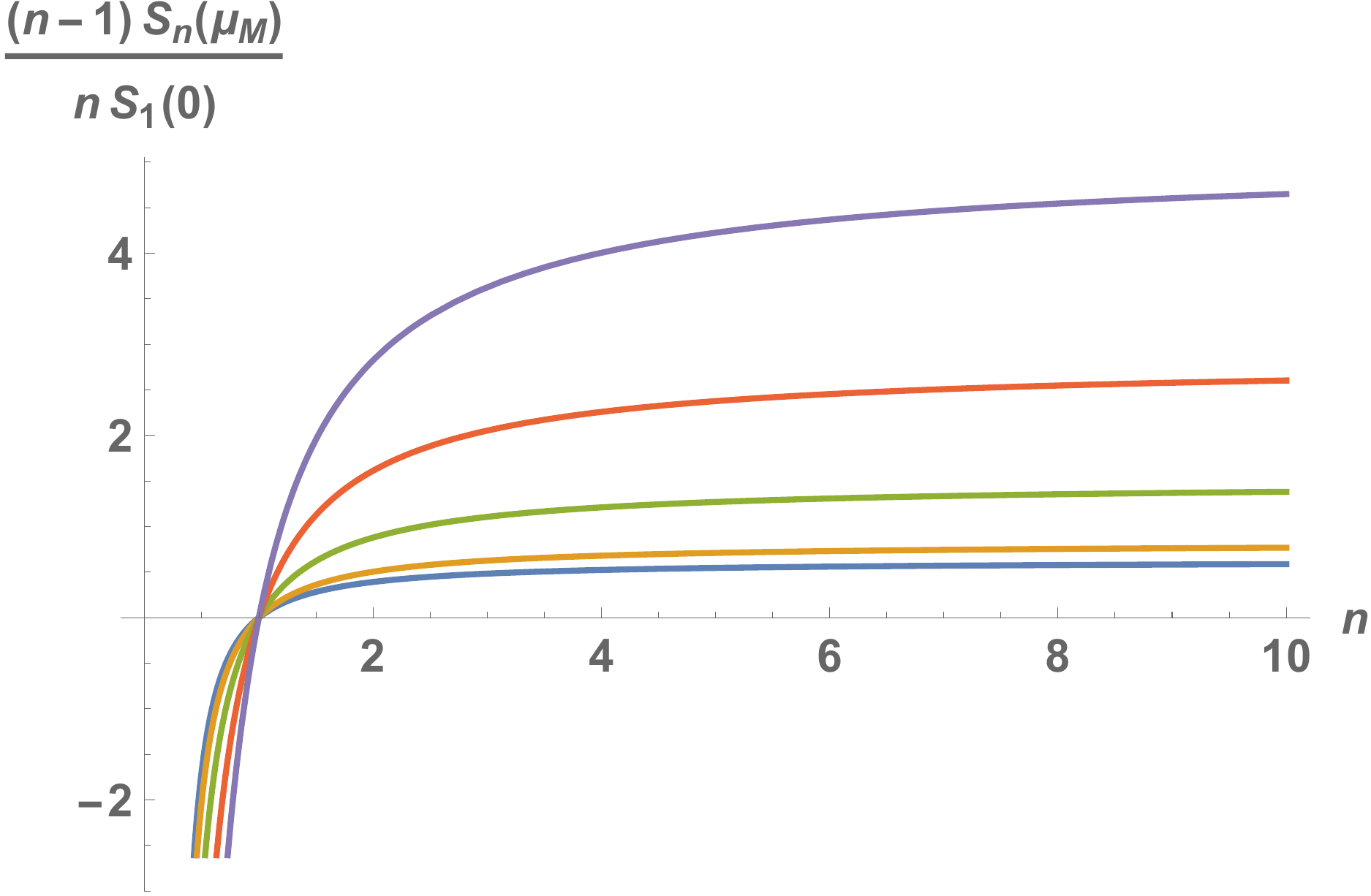}
  \captionof{figure}{Charged Renyi entropy for Maxwell fields in $d=4$ normalized by the entanglement entropy with zero chemical potential on the left panel, and $\frac{n-1}{n}\frac{S_n(\mu)}{S_1(0)}$ on the right panel. The UV cutoff has been chosen such that $\V = 10^{40}$~. The curves corresponds to $\tmu_{\rm M}=0,0.5,1,1.5,2$ from bottom to top.  }
  \label{fig:max1}
  \end{center}
\end{figure}

\begin{figure}[t!]
\begin{center}
  \includegraphics[scale=0.4]{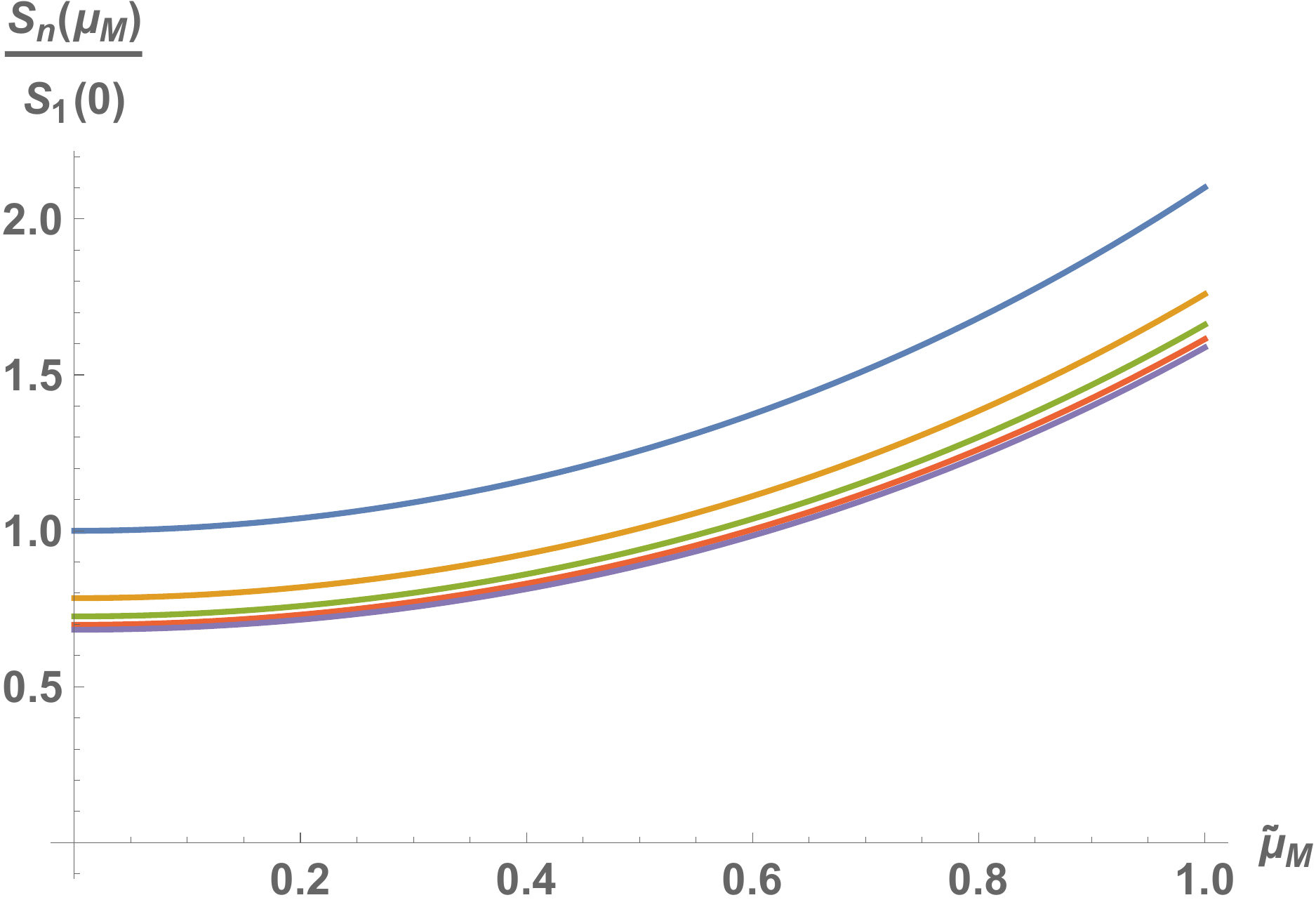}
  \includegraphics[scale=0.4]{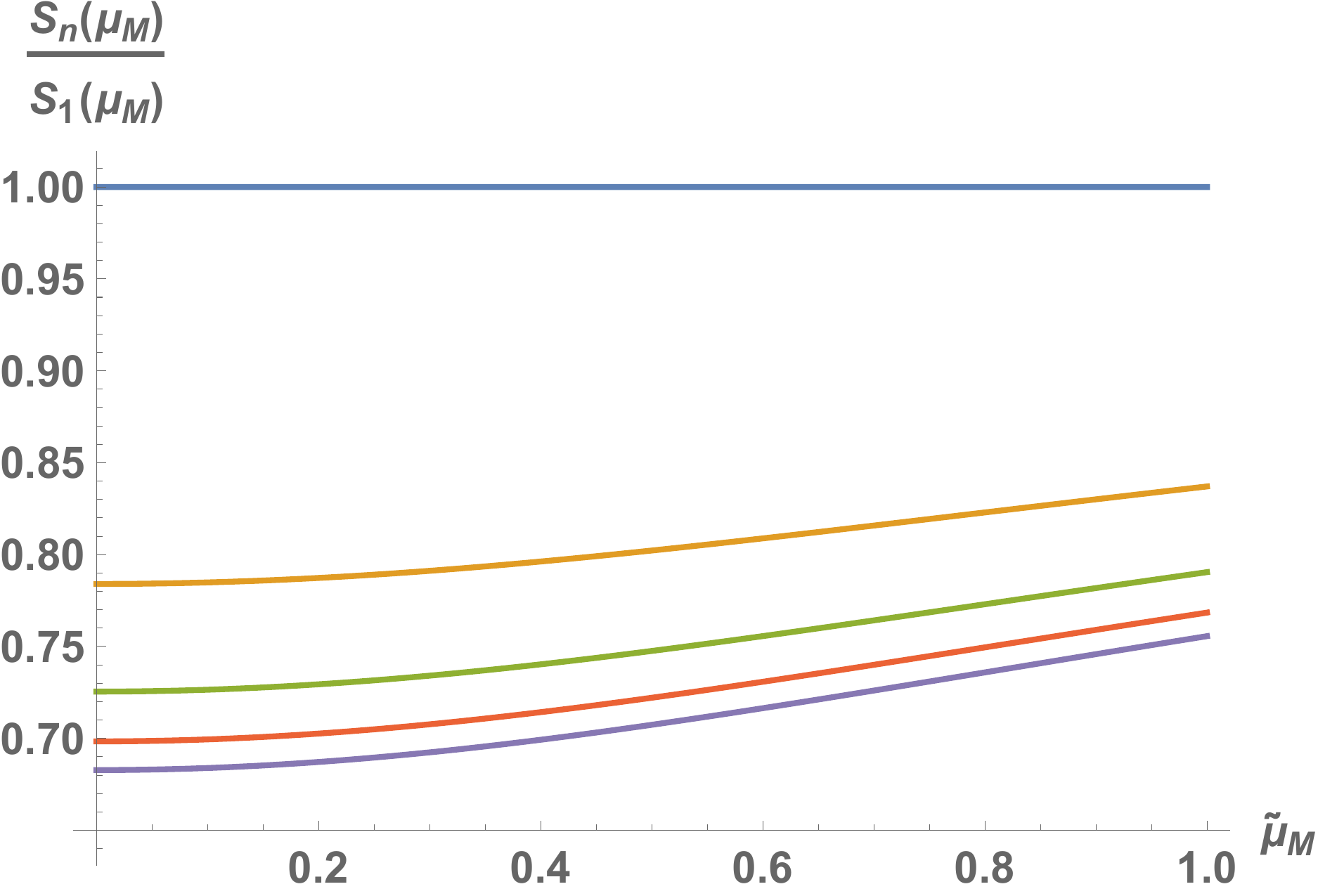}
  \captionof{figure}{Charged Renyi entropy for Maxwell fields in $d=4$ normalized by the entanglement entropy with zero chemical potential on the left panel, and non-zero chemical potential on the right panel. The UV cutoff has been chosen such that $\V = 10^{40}$~. The curves corresponds to $n=1,2,3,4,5$ from top to bottom.  }
  \label{fig:max2}
  \end{center}
\end{figure}

An interesting feature is that, to lading order in $\mu$, the Renyi entropy is $n$ independent \cite{Belin:2013uta} (maximal disorder), and this remains true when the logarithmic corrections are considered
\begin{align}
    \lim_{\mu\to\infty}S_n(\mu) = \V \left(\frac{d-2}{d}\right)^{\frac{d-1}{2}}\tmu_{\rm M}^{d-1}-\nC_{\rm M} \log\left[\V \left(\frac{d-2}{d} \right)^{\frac{d-1}{2}}\tmu_{\rm M}^{d-1}\right]~,
\end{align}
As can be checked by the numerical results plotted in \autoref{fig:max1}, the Renyi inequalities \eqref{ineq} hold when the coefficient $\nC$ is not comparable with the horizon volume. As can be seen from \autoref{fig:max2}, the growth in the entangled pairs created with the increasing of real chemical potential holds also as in \cite{Belin:2013uta}. 

Finally, the conformal weights of the dual twist operators are 
\begin{align}
    h_n(\mu) ={}& \pi n \left(\frac{L}{\lp}\right)^{d-1}\left( x_n^{d-2} - x_n^{d} \right) - \frac{n\nC}{2V_{\mathbb{H}^{d-1}}}\frac{(1 - x_n)(d x_n + d-2)}{x_n}- \frac{n x_n^{d-2}}{2\pi R^d}\frac{L^{d-2}}{\lp^{d-1}}\tmu_{\rm M}^2~, \\
    \partial_n h_n\Big\rvert_{n=1,\mu=0} ={}& \frac{2\pi}{d-1}\left(\frac{L}{\lp}\right)^{d-1} - \frac{\nC}{V_{\mathbb{H}^{d-1}}}~,
\end{align}

Then, the extra subleading logarithmic divergence in the entanglement entropy \eqref{SEMax}, which has been discussed in \autoref{Sec:Conclusions} to be related to systems with SSB, also appearin higher-dimensional Maxwell electrodynamics. This remains present in the zero charge limit. 
\newpage



\end{appendix}

\bibliographystyle{JHEP}
\bibliography{biblio}
\end{document}